\RequirePackage{fix-cm}

\documentclass[smallextended]{svjour3}

\usepackage{mathptmx}
\usepackage{psfrag,epsf}
\usepackage{enumerate}
\usepackage{natbib}
\usepackage{hyperref}
\usepackage{amssymb, amsfonts, enumerate}
\usepackage{url}
\usepackage{graphicx}
\usepackage{graphics}
\usepackage{booktabs}
\usepackage{multirow}
\usepackage{hyperref}
\usepackage{color}
\usepackage{bm}
\usepackage{bbm}
\usepackage{parskip}
\usepackage{tikz}
\usepackage[intlimits]{amsmath}
\usepackage{graphicx}
\usepackage{graphics}
\usepackage{booktabs}
\usepackage{hyperref}
\usepackage{color}
\usepackage[margin=1 in]{geometry}
\usepackage[export]{adjustbox}
\usepackage{caption}
\usepackage{subcaption}
\usepackage{setspace}
\usepackage[pagewise]{lineno}

\newcommand{\bx}{\bm{x}}

\newcommand{\btta}{\bm{\theta}}

\journalname{Metrika}

\begin{document}



\title{Statistical Inference Based on a New Weighted Likelihood Approach}

\titlerunning{Statistical Inference Based on a New Weighted Likelihood Approach}

\author{Suman Majumder         \and
        Adhidev Biswas \and 
        Tania Roy \and
        Subir Kumar Bhandari \and
        Ayanendranath Basu
}

\authorrunning{Majumder et al.}

\institute{Ayanendranath Basu \at
              Interdisciplinary Statistical Research Unit, Indian Statistical Institute, 203 Barrackpore Trunk Road, Kolkata, WB 700108, India\\Tel: +91 33 2575 2806\\ \email{ayanbasu@isical.ac.in}\\ ORCID: https://orcid.org/0000-0003-1416-9109\\ Corresponding Author
              \and
           Suman Majumder \at
              Department of Statistics, North Carolina State University, 2311 Stinson Drive, Raleigh, NC 27695, USA\\\email{smajumd2@ncsu.edu}\\ ORCID: https://orcid.org/0000-0002-3678-5642       
           	  \and
           Adhidev Biswas \at
              Interdisciplinary Statistical Research Unit, Indian Statistical Institute, 203 Barrackpore Trunk Road, Kolkata, WB 700108, India\\\email{adhidevbiswas@gmail.com}
              \and
           Tania Roy \at
	          Department of Statistics and Applied Probability, UC Santa Barbara, Santa Barbara, California 93106, USA\\\email{tania112358@gmail.com}
              \and
           Subir Kumar Bhandari \at
              Interdisciplinary Statistical Research Unit, Indian Statistical Institute, 203 Barrackpore Trunk Road, Kolkata, WB 700108, India\\\email{subir@isical.ac.in}
}

\date{Received: date / Accepted: date}

\maketitle
\begin{abstract}
We discuss a new weighted likelihood method for parametric estimation. The method is motivated by the need for generating a simple estimation strategy which provides a robust solution that is simultaneously fully efficient when the model is correctly specified. This is achieved by appropriately weighting the score function at each observation in the maximum likelihood score equation. The weight function determines the compatibility of each observation with the model in relation to the remaining observations and applies a downweighting only if it is necessary, rather than automatically downweighting a proportion of the observations all the time. This allows the estimators to retain full asymptotic efficiency at the model. We establish all the theoretical properties of the proposed estimators and substantiate the theory developed through simulation and real data examples. Our approach provides an alternative to the weighted likelihood method of \citet{markatou97,markatou98}.


\end{abstract}


\section{Introduction} \label{sec: 1}
	We consider a new approach to weighted likelihood estimation. A weighted likelihood estimating equation employs a reweighting of the components of the likelihood score equation. This method is useful when the model is in doubt or when outliers are present in the data. The weighted likelihood estimator considered here (obtained as a solution of the weighted likelihood estimating equation) is asymptotically fully efficient in cases where the model is true, and in cases where the model is perturbed the proposed estimator works robustly, identifying the points in the data that are not in agreement with the model.
	
	The method we discuss is based on a recent proposal by \cite{biswas14}. This work simply puts forward the proposal in a brief article; detailed numerical investigations or derivation of the theoretical properties of the method have not been undertaken. In the present paper we provide a comprehensive follow-up of the proposal, derive its theoretical properties, describe the possible extension of the method to situations beyond the independent and identically distributed (i.i.d.) univariate data model and consider extensive numerical explorations; overall we provide a general discussion of the scope of the application of the method in statistical inference.
	
	Let ${X}_1,{X}_2,\cdots,{X}_n$ be an i.i.d. random sample from a distribution $G$ having density $g$. We model $G$ by the parametric family $\mathcal{F}_\Theta=\{ F_\theta : \theta \in \Theta \subset \mathbb{R}^d \}$. Let $u_{\theta}(x)=\nabla \log[f_{\theta}(x)]$ be the likelihood score function where $f_{\theta}$ is the density corresponding to $F_{\theta}$, $\nabla$ denotes differentiation with respect to $\theta$. Under usual regularity conditions, the maximum likelihood estimator (MLE) of $\theta$ is obtained as a solution to the score equation $\sum_{i=1}^n u_{\theta}({X_i})=0$.
	
	For any given point $t$ in the sample space, we construct a weight function $ w_\theta(t) \equiv H(t,F_{\theta},F_n) $ that depends on the point $t$, the model $ F_\theta $ and the empirical distribution function $ F_n $. The weights will be constrained to lie between $ 0 $ and $ 1 $. Ideally, the weights should be close to $ 1 $ for points where the data closely follow the model but should be substantially smaller when the two do not agree. The weighted likelihood estimator (WLE) of $\theta$ is the solution of
	\begin{equation} \label{eq:WLEE}
		\sum_{i=1}^n w_\theta(X_i) u_\theta(X_i) = 0.
	\end{equation}
	Here $ w_\theta(X_i) \equiv H(X_i,F_\theta,F_n) $ is the weight attached to the score function of $X_i$. The function $ H(\cdot) $ is independent of $ \theta $, and the dependency of $ \theta $ in $ w_\theta(\cdot) $ comes solely from $F_\theta$, a component of the argument of $ H(\cdot) $. To be consistent with our philosophy, the weights should be equal to $1$ at the points where $F_\theta$ and $F_n$ are identical. The weights should go down smoothly as $F_n$ becomes more disparate with $F_\theta$ in either direction.

	One of the first approaches to weighted likelihood estimation based on discrepancies between the data and the assumed model was presented by \cite{green1984iteratively}. This was further refined by \cite{lenth1987consistency} who updated the discrepancy function using Huber's $\psi$ function \citep{huber1964robust}. \cite{field1994robust} proposed to downweight observations lying outside the central $100\times(1-2p)\%$ of the distribution, $0 < p \leqslant 1/2$. \cite{markatou97,markatou98} considered a weighted likelihood approach to estimation based on weights quantifying the magnitude and sign of the Pearson residual linked to a residual adjustment function employed in minimum disparity estimation; see \cite{lindsay94} and \cite{basu94}. The pioneering works of \cite{markatou97,markatou98} provide a useful procedure that simplifies the estimation technique of minimum disparity estimation, specially in continuous models. In particular, the estimating function is reduced to a sum over the observed data rather than an integral over the entire support. \citet{gervini2002class} represent another important work in this spirit. 

	The present work provides an alternative to the \cite{markatou97,markatou98} procedure with an aim to further refine the technique, without compromising its good qualities. The weighted likelihood methods in this branch of inference depend on two different quantities, the residual function, and the weight function. The former quantifies the compatibility between the data and the model at each observed value; depending on the value of the residual,  the weight function decides the importance of the observation in the estimating equation. Here we provide different approaches for both compared to the \cite{markatou97,markatou98} approach. We compute residuals by comparing the empirical distribution function and the model distribution function, rather than comparing a nonparametric kernel density estimate of the true distribution with the probability density function (PDF) of the model. This avoids the possible difficulties associated with kernel density estimation, such as bandwidth selection, slow rate of convergence and problems of bounded support. We also generalize the selection of weights by choosing functions that satisfy some basic criteria, rather than choosing only such weights that are linked to the residual adjustment function. A more detailed discussion about the advantages of our methods is presented in \href{ESM1.pdf}{Online Resource 1}.
	
	Claudio Agostinelli has extensively studied the form of the weighted likelihood estimators proposed by \cite{markatou98} and applied them to different inference scenarios and generated useful robust estimators and other inference tools. See, for example, \cite{agostinelli1998one,agostinelli2001test}; \cite{agostinelli2002arobust,agostinelli2002brobust,agostinelli2007robust}; \cite{agostinelli2013weighted}. \cite{agostinelli2018weighted} have addressed the problem of extending the idea of weighted likelihood estimation to multivariate data with elliptical structures. In all of these cases our residual and weight functions will provide an alternative approach to the corresponding inference problem.
	
	The rest of the paper is organized as follows. In Section~\ref{sec: residual and weight}, we describe the residual function and propose several weight functions. These weight functions represent a rich collection of shapes and allow many other possibilities in comparison to the simplistic squared exponential weight function considered in \cite{biswas14}. Neither are they restricted to the weight functions generated by the residual adjustment function. In Section~\ref{sec: Examples}, we illustrate the performance of the estimation method through real data examples. A relevant simulation study is provided in Section~\ref{sec: simulation}. The theoretical properties of the estimator are discussed in Section~\ref{sec: properties}. As the first order influence function turns out to be a poor descriptor of the robustness of our estimators, we take up a higher order influence function analysis of these weight functions in Section~\ref{sec: hif}. The method is applied to bivariate data and robust regression problems in Section~\ref{sec: extension}. Concluding remarks are given in Section~\ref{sec: conclusion}.

\section{The Residual Function and the Weight Function} \label{sec: residual and weight}
	\subsection{The Residual Function} \label{subsec: residual}
		Let $ \mathbf{1}_A $ represent the indicator function of the event $ A $. We define $ F_n $ and $ S_n $ as \[ F_n(x) = \frac{1}{n} \sum_{i=1}^n \mathbf{1}_{\lbrace X_i \leqslant x\rbrace}, \,\ S_n(x)=\frac{1}{n}\sum_{i=1}^n \mathbf{1}_{\lbrace X_i \geqslant x\rbrace}.\] These represent the empirical distribution function and the empirical survival function of the data. Let $F_\theta (x) = \mathbb{P}_\theta (X \leqslant x) \text{ and } S_\theta (x) = \mathbb{P}_\theta (X \geqslant x)$ be the corresponding theoretical quantities. Now, the residual function as proposed by \cite{biswas14} can be described through the following steps.
		\begin{enumerate}
			\item Choose a suitable fraction $p \leqslant 0.5$. This tuning parameter will determine the proportion of observations on either tail that will be subjected to (possible) downweighting.
			\item Assign weights $w_\theta({X}_i)=1$ to all observations satisfying $p < F_\theta({X}_i) < 1-p$. Thus observations belonging to the central $100 \times (1 - 2p)\%$ of the distribution at the current estimated value will get weights equal to 1.
			\item For each value ${X}_i$ in the lower tail, i.e. with $F_\theta({X}_i) \leqslant p$, we consider a possible downweighting of the observation as follows. Compute \[\tau_{n,\theta}({X}_i)=\frac{F_n({X}_i)}{F_\theta ({X}_i)} - 1\] which may be viewed as a standardized residual in comparing the two distribution functions. If the two values $F_n({X}_i)$ and $F_\theta({X}_i)$ are severely mismatched, i.e. the value of $\tau_{n,\theta}({X}_i)$ differs substantially from 0 (in either direction), we treat it as a case that requires downweighting.
			\item For each value ${X}_i$ in the upper tail, i.e., with $F_\theta({X}_i) \geqslant 1-p$, we consider a similar downweighting, but in this case we construct the standardized residuals as\[\tau_{n,\theta}({X}_i)=\frac{S_n({X}_i)}{S_\theta({X}_i)} - 1.\]
		\end{enumerate}
		Thus the final form of the residual function $\tau_{n,\theta}({X}_i)$ is 
		\begin{equation}\label{eq:ResidualFunction}
			\tau_{n,\theta}({X}_i)=
			\left\{ 
			\begin{array}{cl}
				\frac{F_n({X}_i)}{F_\theta({X}_i)} - 1,& \text{if } 0 < F_\theta({X}_i) \leqslant p,\\		
				0,& \text{if } p < F_\theta({X}_i) <1-p,\\
				\frac{S_n({X}_i)}{S_\theta({X}_i)} - 1,& \text{if } 1-p \leqslant F_\theta({X}_i) < 1.\\	
			\end{array}
			\right.
		\end{equation}
		The rationale for defining the residual function in the above manner has been described in \cite{biswas14}. It is clear that the consideration of the distribution function in the left tail and the survival function in the right tail helps highlight the mismatch of the data and the model in the respective tails.
	\subsection{The Weight Function} \label{subsec: weight}
		Given the residual function $\tau_{n,\theta}(X_i)$, the weight given to this $X_i$ based on $n$ i.i.d. observations is denoted by $ w_\theta(X_i) = H(\tau_{n,\theta}({X}_i)) $. To adhere to our requirements, the weight function should have the following properties.
		\begin{enumerate}
			\item $ H(\cdot) $ is a smooth function on $ [-1, \infty) $ with $ 0 \leqslant H(x) \leqslant 1 $, $ H(0) = 1 $ and $ \underset{x\rightarrow \infty}{\lim} H(x) = 0 $.
			\item $ H(-1) $ is small, preferably close to zero; at any rate, substantially smaller than 1.
			\item $ H $ admits two derivatives at zero with $ H^\prime(0) = 0 $ and $ H^{\prime\prime}(0) < 0 $.
		\end{enumerate}
		To determine a possible weight function having the above mentioned properties, we proceed as follows.
		\begin{enumerate}
			\item First we find a nonnegative function $ g_{\gamma}(x) $ from $ [a,\infty) \rightarrow \mathbb{R} $ for each fixed value of $\gamma$ and some $ a \in \mathbb{R} $. Suppose for any fixed $\gamma$, $g_\gamma (a)=0$. Moreover, let the function $g_\gamma (x)$ have a unique mode in an interior point of the interval $[a,\infty )$, $g_\gamma (x)$ is bounded, and $g_\gamma (x)$ is twice differentiable in $x$ for each fixed value of $\gamma$.
			\item Then, if possible, we choose the parameter values such that, if the domain is $[a, \infty )$, the unique mode is at $(a+1)$. Let this function be $g_{\gamma_0}$, where $\gamma_0$ provides the required parametrization.
			\item We define the weight function as
			\begin{equation}
				H_{\gamma_0}(\tau_{n,\theta}(x)) = \frac{g_{\gamma_0}(\tau_{n,\theta}(x)+a+1)}{g_{\gamma_0}(a+1)}.
			\end{equation}
			This procedure gives candidates for weight function such that $ H(-1) = 0,~H(0) = 1 $ and $ H^\prime(0)=0$. Using this procedure, several weight functions are proposed.
		\end{enumerate}
		\subsubsection{Weight Functions 1, 2, 3, 4} \label{subsubsec: W1}
\begin{figure}[!htbp]
		\centering
		\begin{subfigure}{0.5\textwidth}
			\centering
			\includegraphics[width=\textwidth, height=0.85\textwidth]{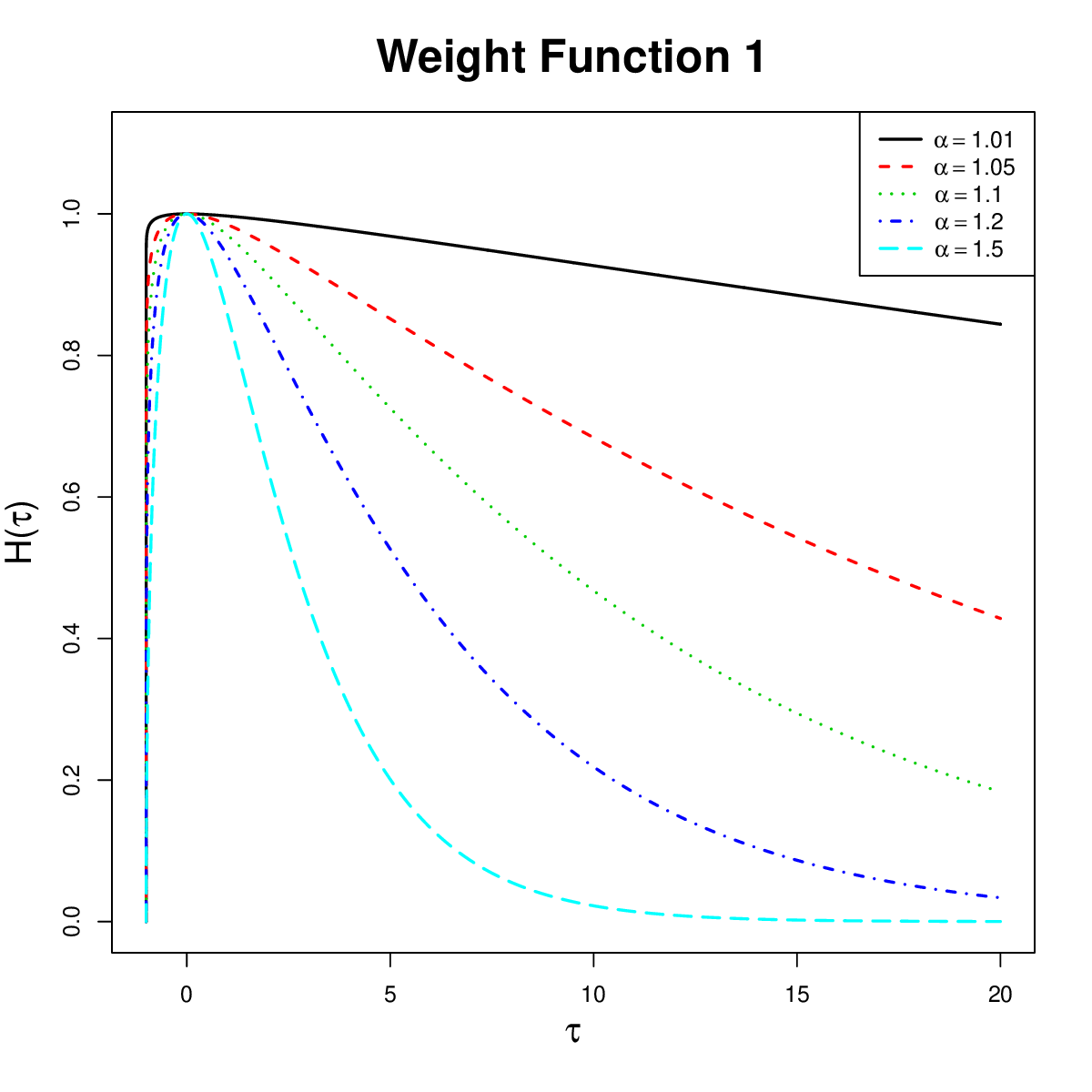}
			\subcaption{Weight function 1} 
			\label{fig: gamma}
		\end{subfigure}%
		\begin{subfigure}{0.5\textwidth}
			\centering
			\includegraphics[width=\textwidth, height=0.85\textwidth]{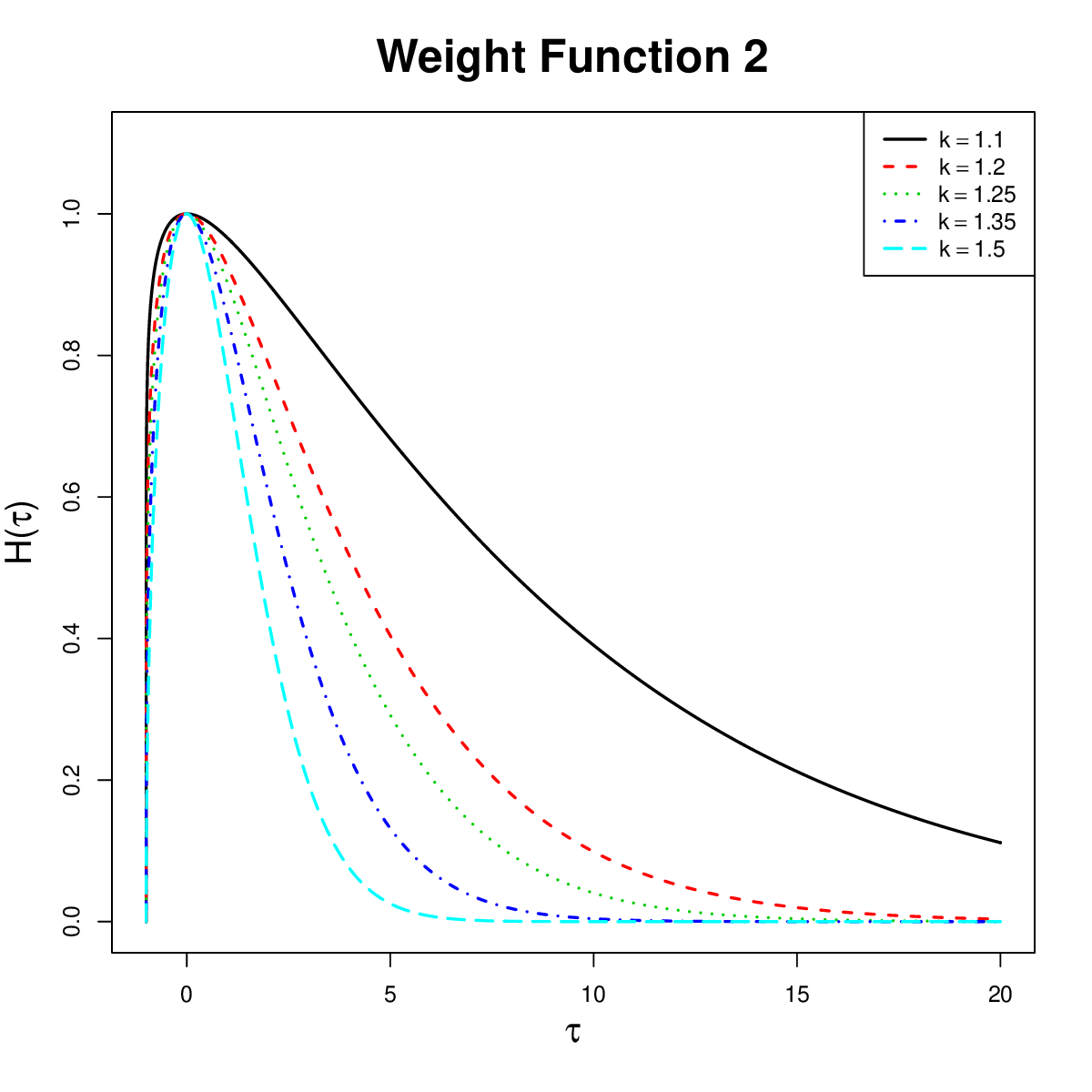}
			\subcaption{Weight function 2}
			\label{fig: weibull}
		\end{subfigure}
	\caption{The shapes of weight functions 1 and 2 for different values of their tuning parameters $\alpha$ and $k$ respectively}
	\end{figure}

			We take $g_\gamma(x)$ as the PDF of a gamma random variable with scale parameter $\lambda$ and shape parameter $\alpha$, which has the form
			\begin{equation*}
				g_{\left(\lambda,\alpha \right)}(x) = \frac{1}{\lambda^{\alpha}\Gamma(\alpha)}x^{(\alpha - 1)}\exp\left(-\frac{x}{\lambda}\right)\,,\ x>0.
			\end{equation*}
			It is bounded, twice differentiable, has a unique mode and the domain is [0,$\infty$). The parametrization $ \lambda= \frac{1}{\alpha - 1}, \alpha > 1 $ sets the mode at $ 1 $. So, the final weight function is $ H_\alpha(\tau_{n,\theta}(x)) = \frac{ g_{\left(1/(\alpha - 1),\alpha \right)}(\tau_{n,\theta}(x)+1)}{g_{\left(1/(\alpha - 1),\alpha \right)}(1)} $.
			As $\alpha$ increases, so does the amount of downweighting and as $\alpha \downarrow 1$ the weights tend to 1 for all values of $\tau_{n,\theta}(x)$, and will eventually lead to the original score equation $ \frac{1}{n}\sum_{i=1}^nu_\theta(X_i) = 0 $, which yields the MLE for the data. Figure~\ref{fig: gamma} presents the shapes of this weight function for different values of the tuning parameter $\alpha$. We denote this weight as weight function 1.
		
			Next, we take $g_\gamma(x)$ to be the PDF of a Weibull distribution with scale parameter $\lambda$ and shape parameter $k$, which has the form
			\begin{equation*}
				g_{(k,\lambda)}(x) =\frac{k}{\lambda}\left( \frac{x}{\lambda} \right)^{(k-1)}\exp\left[-\left(\frac{x}{\lambda} \right)^k\right],~x>0.
			\end{equation*}
			This function is bounded, twice differentiable and takes values in the interval [0,$\infty $) and has a unique interior mode. The parametrization $ k > 1 $ and $ \lambda = \left( \frac{k-1}{k} \right)^{-1/k} $ fixes the mode at 1. The final weight function is $ H_k(\tau_{n,\theta}(x)) = \frac{g_{\left(k,\left(1-1/k\right)^{-1/k}\right)}(\tau_{n,\theta}(x)+1)}{g_{\left(k,(1-1/k)^{-1/k}\right)}(1)} $.
			Downweighting increases with $k$, and the weights converge to $1$ as $k \downarrow 1$. The pattern of the weights for different choices of $k$ is displayed in Figure~\ref{fig: weibull}. We denote this as weight function 2.
		
\begin{figure}[!htbp]
		\centering
		\includegraphics[width=0.8\linewidth, height=0.45\linewidth]{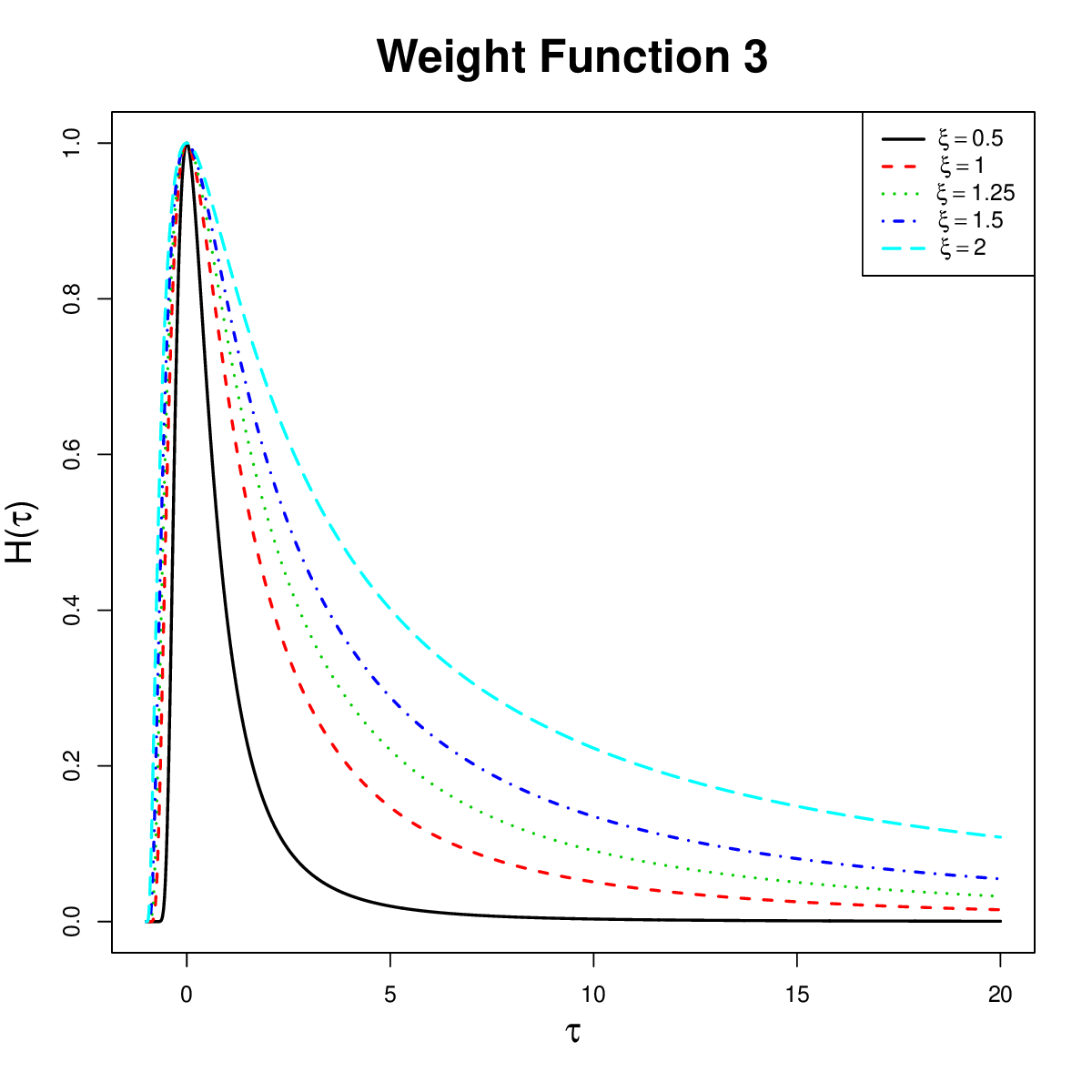}
		\caption{The shapes of weight function 3 for different values of the tuning parameter $\xi$}
		\label{fig: gev}
	\end{figure}

			Now we take our $g_\gamma(x)$ to be the PDF of the generalized extreme value (GEV) distribution with location parameter $\mu$, scale parameter $\beta$ and shape parameter $\xi$ which has the form
			\begin{equation*}
				g_{(\mu,\beta,\xi)}(x)=\frac{1}{\beta}t(x)^{(1+\xi)}\exp(-t(x)),~\text{where}~~t(x)=\left(1+ \left(\frac{x - \mu}{\beta}\right)\xi\right)^{-1/\xi}.
			\end{equation*}
			To set the starting point of the domain at $-1$ and the mode at 0, we reparametrize the function as $\xi > 0$, $\mu=(1 + \xi)^\xi - 1$ and $\beta=\xi (1 + \xi)^\xi$. So, the final weight function is
			\begin{equation*}
				H_\xi(\tau_{n,\theta}(x)) = \frac{g_{((1 + \xi)^\xi - 1,\xi (1 + \xi)^\xi,\xi)}(\tau_{n,\theta}(x))}{g_{((1 + \xi)^\xi - 1,\xi (1 + \xi)^\xi,\xi)}(0)}.	
			\end{equation*}
			Downweighting decreases with increasing $\xi$ (see Figure~\ref{fig: gev}). This is our weight function 3.

\begin{figure}[!htbp]
		\centering
		\begin{subfigure}{0.5\textwidth}
			\centering
			\includegraphics[width=\textwidth, height=0.95\textwidth]{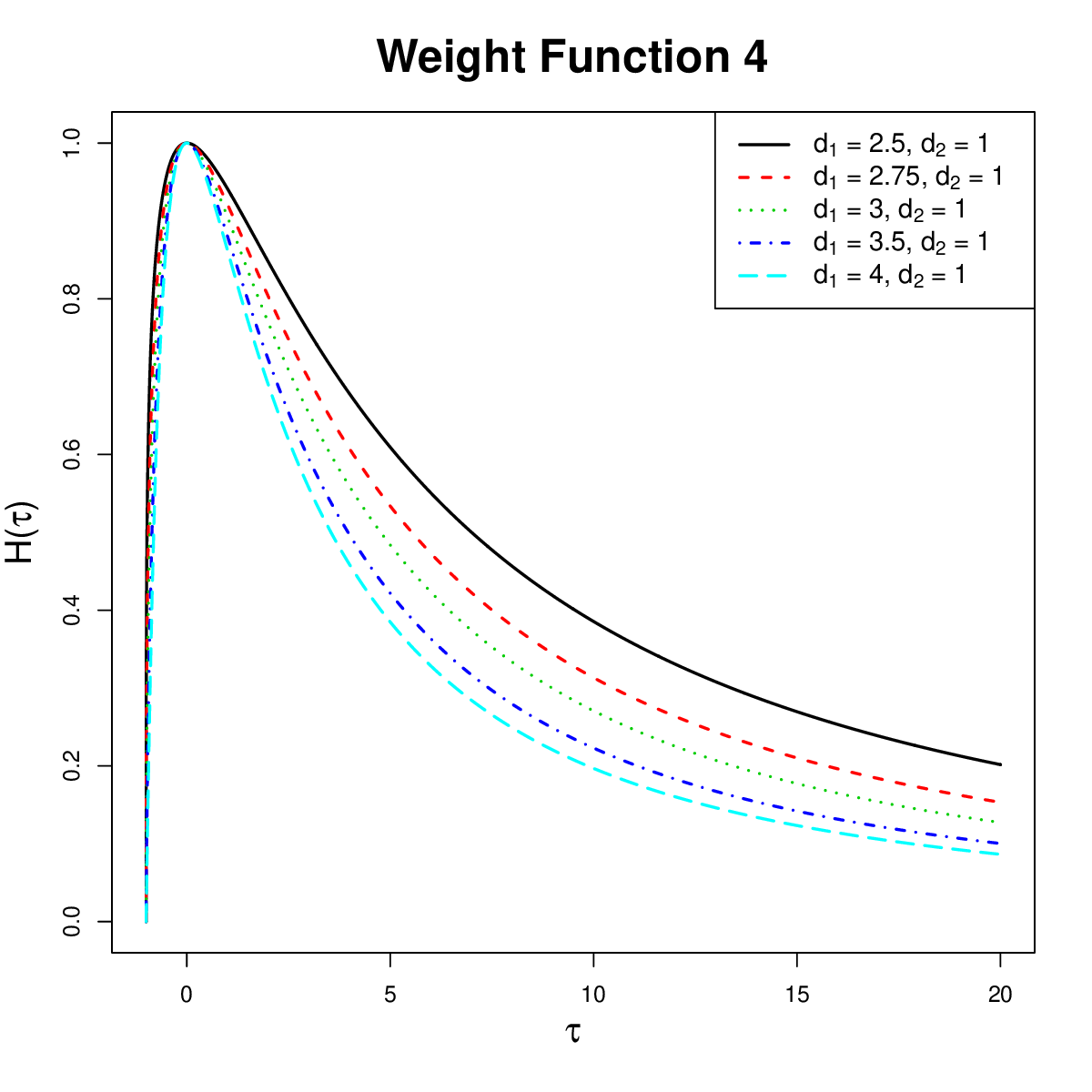}
			\subcaption{Tuning $d_1$}
			\label{fig: f1}
		\end{subfigure}%
		\begin{subfigure}{0.5\textwidth}
			\centering
			\includegraphics[width=\textwidth, height=0.95\textwidth]{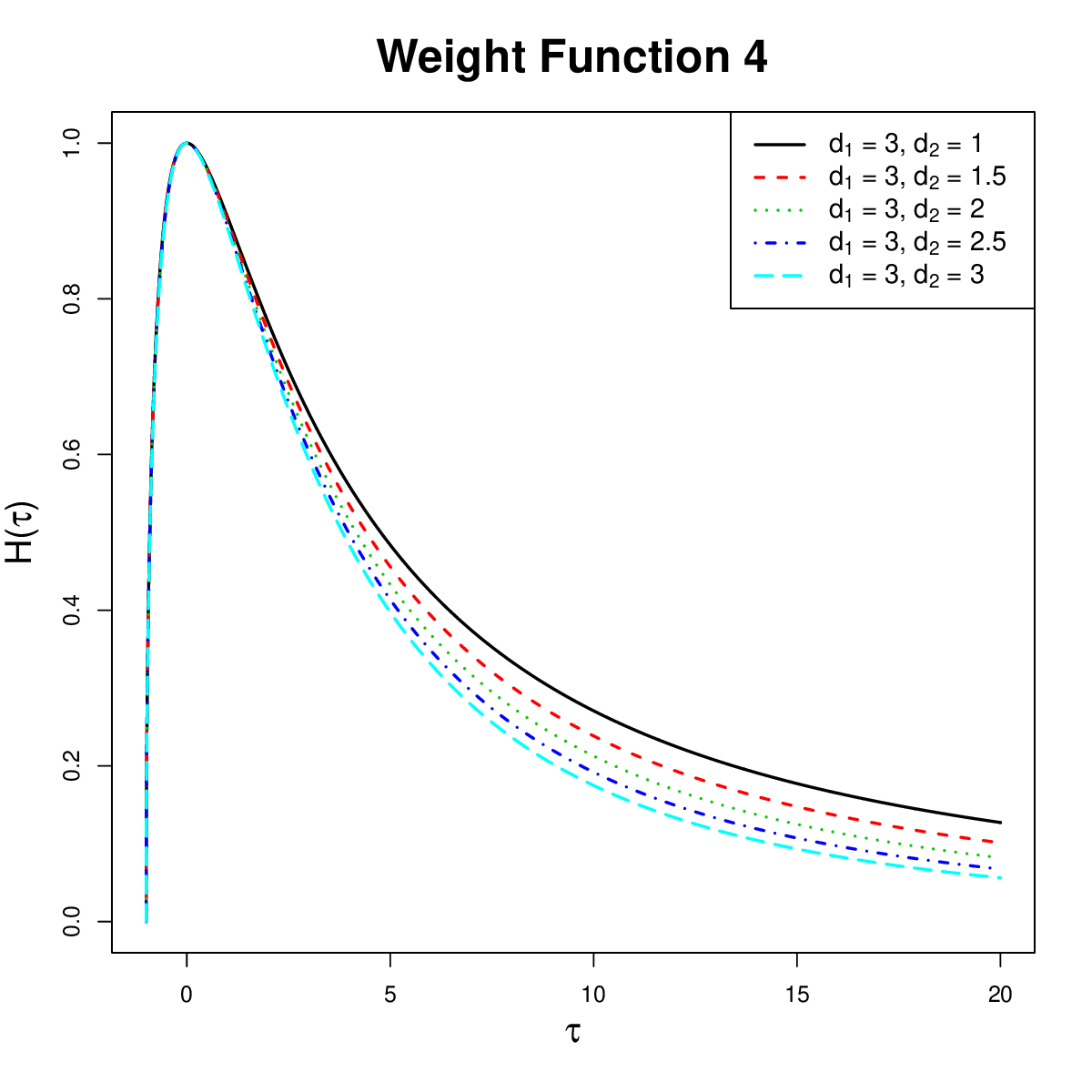}
			\subcaption{Tuning $d_2$}
			\label{fig: f2}
		\end{subfigure}
	\caption{The shapes of weight function 4 for tuning $d_1$ and $d_2$ respectively, keeping the other fixed}
	\end{figure}

			We now take $ g_\gamma(x)$ to be the function
			\begin{equation*}
				g_{(d_1,d_2,a)}(x) = \frac{1}{aB(d_1/2,d_2/2)} \left(\frac{d_1}{d_2}\right)^{d_1/2}\left(\frac{x}{a}\right)^{\left(d_1/2-1\right)}\left(1 + \frac{d_1x}{d_2a}\right)^{-(d_1+d_2)/2}
			\end{equation*}
			on the domain [0,$\infty$), where the parameters $d_1,d_2,a$ are all positive and $ B(\cdot, \cdot) $ is the beta function. The function is integrable and normalized. To fix the unique interior mode at 1, we parametrize the function as $a=\frac{d_1(d_2+2)}{(d_1-2)d_2}$, $d_1>2$, $d_2>0$. So, the final weight function is given by
			\begin{equation*}
				H_{d_1,d_2}(\tau_{n,\theta}(x)) = \frac{g_{\left(d_1,d_2,\frac{d_1(d_2+2)}{(d_1-2)d_2}\right)}(\tau_{n,\theta}(x)+1)}{g_{\left(d_1,d_2,\frac{d_1(d_2+2)}{(d_1-2)d_2}\right)}(1)}.
			\end{equation*}
			As $d_1 \downarrow 2$, the weights tend to 1 and the weighted likelihood equation tends to the maximum likelihood score equation. The varying weight functions with $d_2$ fixed at 1, and $d_1$ fixed at 3 are depicted in Figures~\ref{fig: f1} and \ref{fig: f2} respectively. The second tuning parameter $d_2$ has no particular effect on the left tail when the other tuning parameter $d_1$ is kept constant. The parameter $d_1$, however, has an effect on both the tails. Thus one can control the right tail of the function without affecting the left tail. This is weight function 4.
		\subsubsection{Choice of Weight Function} \label{subsubsec: weight_choice}
			The conditions on our weight function, as presented in Section \ref{subsec: weight}, are rather general and as a result, one can come up with many a weight function that are suitable for our purpose. This article itself proposes four classes of such weight functions. This creates a conundrum in a practitioner's mind as to what choice of weight function should be made for which problem. Does it matter which weight is chosen and if so, then, how does it affect the outcome? These questions are natural and need to be addressed.

			When the model is correctly specified, the choice of the weight functions does not affect the important theoretical properties of the estimator such as influence function, location-scale equivariance, consistency or asymptotic efficiency, as seen in Section \ref{sec: properties}. There are mild conditions attached to the weight functions being used in these scenarios which are all automatically satisfied by construction of these weights. In fact, all the theoretical properties are intact for any choice of tuning parameter(s) for these weight functions.

			In an applied sense, it is more important to get a proper downweighting of the outliers. For all of the weight functions proposed in this article, any specified level of downweighting at any specified level of the residual can be achieved by properly tuning our weight functions. This is not possible with the fixed cohort of standard disparity weights. However, given that each of our four classes of weights have a rich and extensive collection of shapes, it will take more research to choose between them. In the present work we have, unless otherwise mentioned, worked with weight function 1 (and found it quite satisfactory). Also, wherever necessary, the tuning parameters $ \alpha $, $ k $, $ \xi $ and $ (d_1, d_2) $ are used to distinguish between the different weight functions.
\section{Some Real Data Examples} \label{sec: Examples}
	\subsection{Drosophila Data} \label{subsec: drosophila}
		The experimental protocols for this chemical mutagenicity data set are available in \citet{woodruff84}. In this experiment which involved drosophila, a variety of fruit flies, the experimenter exposed groups of male flies to different doses of a chemical to be screened. Subsequently each male was mated with unexposed females. Approximately 100 daughter flies were sampled for each male and the experimenter noted the number of daughters carrying a recessive lethal mutation on the X chromosome. The data set consisted of the observed frequencies of males having 0, 1, 2, $\cdots$ recessive lethal daughters. Data for the specific experimental run on day $177$ are presented in Table~\ref{tab: droso-data}.
		\begin{table}[!htbp]
			\begin{center}
				\caption{Drosophila data}
				\begin{tabular}{c c c c c c c}
					\hline
					No. of daughters&$0$&$1$&$2$&$3$&$4$&$\geqslant 5$\\
					Observed frequency&$23$&$7$&$3$&$0$&$0$&$1(91)$\\
					\hline
				\end{tabular}
				\label{tab: droso-data}
			\end{center}
		\end{table}
		\begin{table}[!htbp]
			\begin{center}
				\caption{Parametric estimates obtained for the drosophila data using different methods}
				\begin{tabular}{c c c c c c c c}
					\hline
					Method&MLE&MLE-D&HD&GKL$_{1/3}$&RLD$_{1/3}$&WLE$_{1}(\alpha = 1.01)$&WLE$_2(k=1.01)$\\
					$\hat{\theta}$&$3.0588$&$0.3939$&$0.3637$&$0.3813$&$0.3588$&$0.3948$&$0.3948$\\
					\hline
				\end{tabular}
				\label{tab: droso-comp}
			\end{center}
		\end{table}
		A Poisson($\theta$) model was fitted to these data. We set our parameter value for $p=0.5$ for this purpose. We use the same value of $ p $ henceforth, unless otherwise mentioned. The estimated means for various methods are given in Table~\ref{tab: droso-comp}. MLE-D represents the outlier-deleted MLE. The WLEs successfully provide outlier resistant estimates of $\theta$ giving almost 0 weight to the outlier, unlike the MLE. In fact, the WLEs are seen to be very close to the outlier-deleted MLE. Here HD, GKL$_{1/3}$ and RLD$_{1/3}$ represent the Hellinger distance, the generalized Kullback-Leibler divergence and the robustified likelihood disparity respectively (with indicated tuning parameters) and these estimators are as reported by \cite{basu11}, Table $2.2$. All the robust estimators considered here have values within a very small band, whereas the MLE produces a nonsensical result.
 
	\subsection{Newcomb's Speed of Light Data} \label{subsec: newcomb}
\begin{figure}[!htbp]
		\centering
		\includegraphics[width=0.95\linewidth, height=0.7\linewidth]{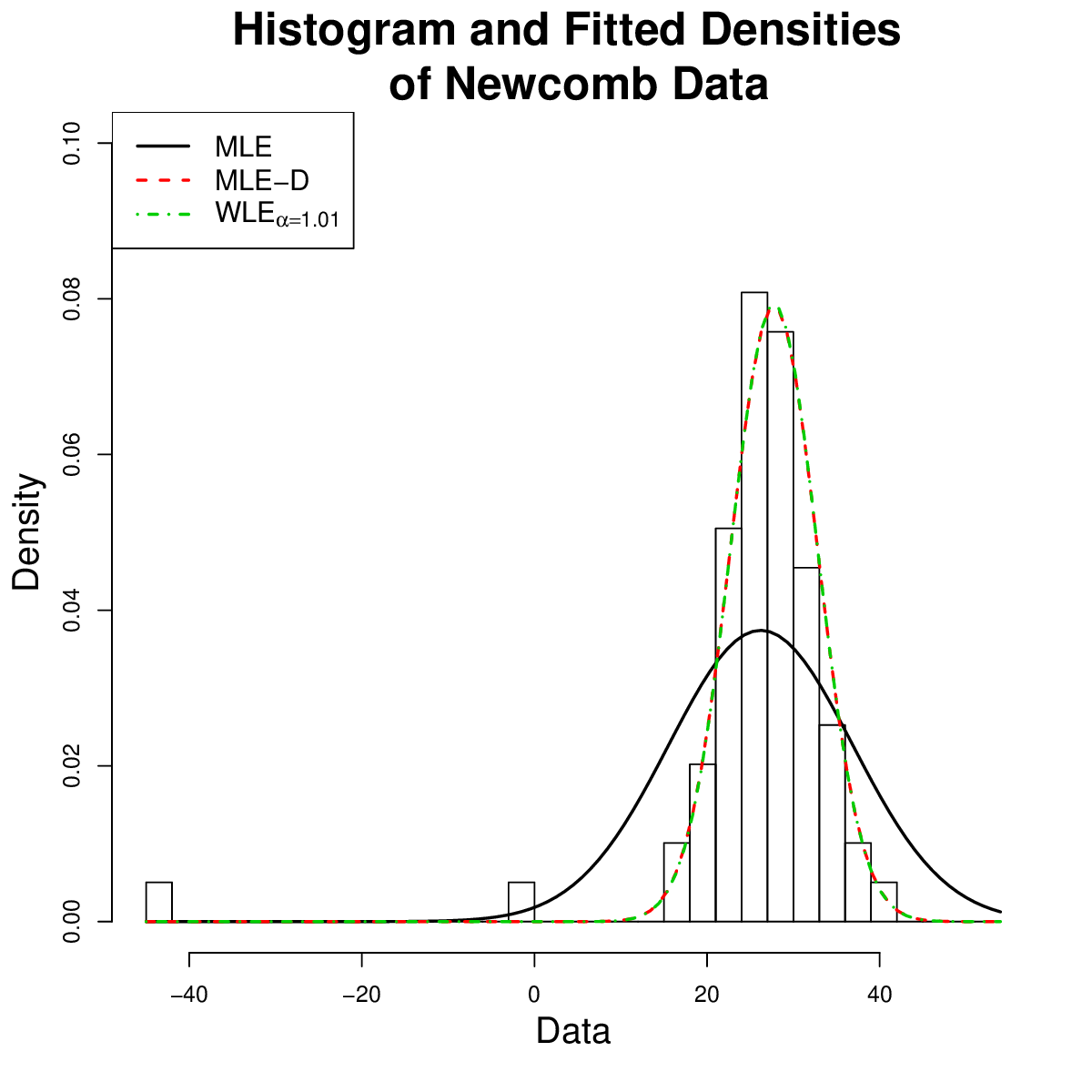} 
		\caption{Histogram and the fitted normal densities to the Newcomb's speed of light data}
		\label{fig: NewcombFits}
	\end{figure}

		Newcomb's speed of light data, available in \cite{stigler77}, consists of $ 66 $ observations. There are two distinct outliers at $-44$ and $-2$. A normal model would provide a nice fit to these outlier-deleted data. The parameter estimates under the normal model are presented in Table~\ref{tab: newcomb-comp} for several WLEs together with their likelihood based competitors. Figure~\ref{fig: NewcombFits} displays the close match of the WLE (with $ \alpha = 1.01 $) with the outlier-deleted MLE. All the WLEs are successful in controlling the effect of the outliers while the MLE is severely affected.
		\begin{table}[!htbp]
			\begin{center}
				\caption{Estimates obtained for Newcomb's speed of light data using different estimators}
				\begin{tabular}{c c c c c c c c c}
					\hline
					Estimator&MLE&MLE-D&&&WLE&&&\\
					\cline{4-9}
					Tuning parameter&-&-&$\alpha = 1.01$&$\alpha = 1.1$&$k=1.05$&$k=1.1$&$\xi=5$&$\xi=10$\\
					\hline
					$\hat{\mu}$&$26.2121$&$ 27.75$&$27.7581$&$27.8460$&$27.7982$&$27.8722$&$27.8303$&$27.7891$\\
					$\hat{\sigma}^2$&$113.7126$&$25.4375$&$25.3204$&$23.9902$&$24.7364$&$23.6171$&$23.7256$&$24.6965$\\
					\hline
				\end{tabular}
				\label{tab: newcomb-comp}
			\end{center}
		\end{table}
	\subsection{Melbourne's Daily Rainfall Data} \label{subsec: rainfall}
\begin{figure}[!htbp]
		\centering
		\includegraphics[width=0.95\linewidth, height=0.7\linewidth]{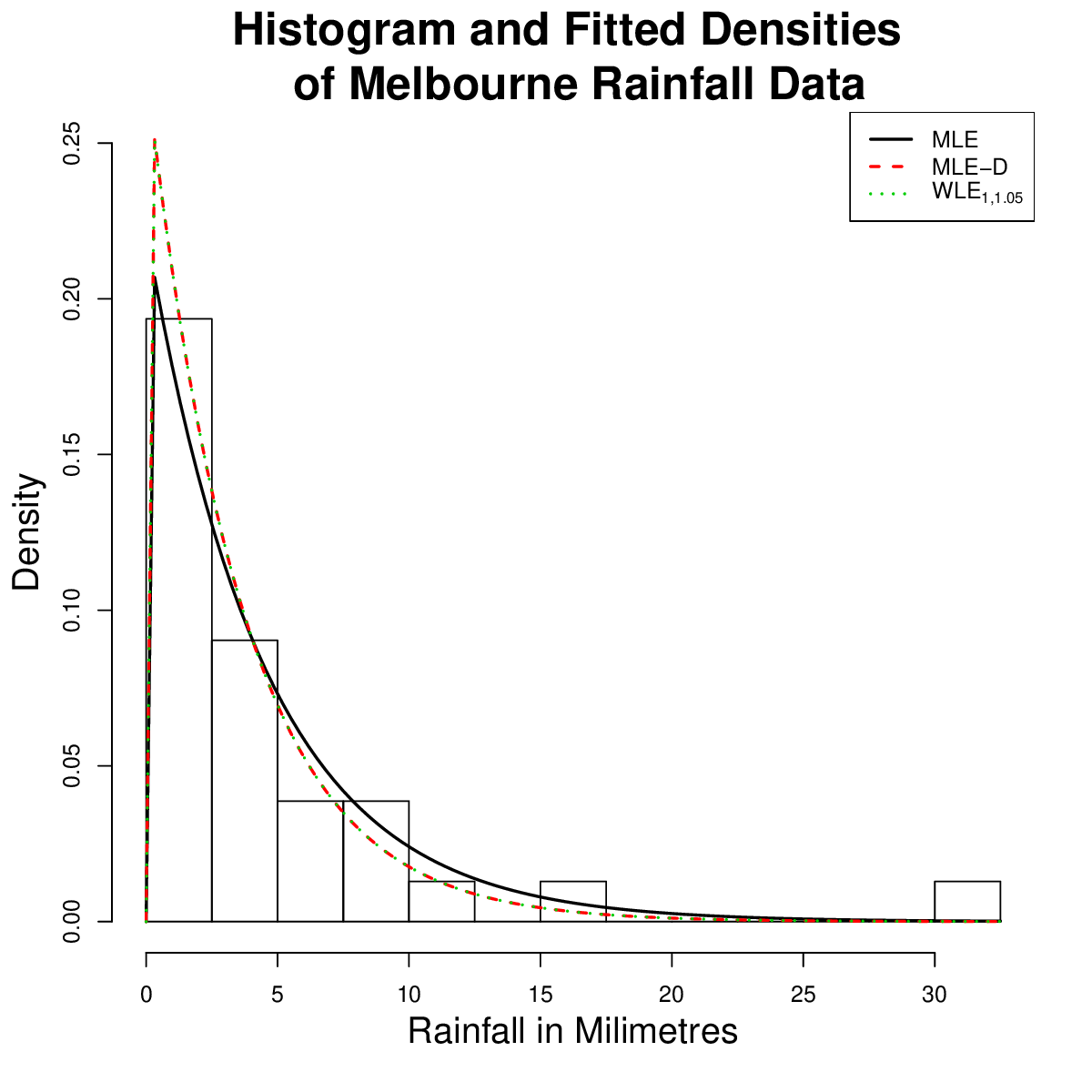} 
		\caption{Histogram and exponential densities fitted to the Melbourne's daily rainfall data}
		\label{fig: rainfall-diagram}
	\end{figure}

		Here we apply our method beyond the domain of the normal model. This data set is on the daily rainfall in Melbourne for the months of June to August in the years 1981 to 1983, presented in \cite{staudte90}. As it is unrealistic to pretend that rainfall over successive days are independent, \citeauthor{staudte90} took the measurements for every fourth rain day, creating a sample of size $31$, and assumed that they are independent and identically distributed. Under this assumption, we fit an exponential model to the data and estimate the rate parameter $\lambda$. Table~\ref{tab: rain-comp} presents the estimates and Figure~\ref{fig: rainfall-diagram} displays the histogram and fitted densities. The proposed WLE does well to overcome the effect of the large outlier in the right tail unlike the MLE.
		\begin{table}[!htbp]
			\begin{center}
				\caption{The estimated parameter values for the Melbourne's daily rainfall data}
				\begin{tabular}{c c c c}
					\hline
					Method&MLE&MLE-D&WLE$_1(\alpha=1.05)$\\
					$\hat{\lambda}$& 0.2224&0.2747&0.2786\\
					\hline
				\end{tabular}
				\label{tab: rain-comp}
			\end{center}
		\end{table}
\section{Simulation Study} \label{sec: simulation}
	In this section, we will present the results of an extensive simulation study to numerically demonstrate the performance of the proposed weighted likelihood estimators in providing high efficiency simultaneously with strong robustness. As the greatest benefit of this method compared to the disparity based methods of inference is in the continuous model, we choose the normal and exponential models for illustration.

	We shall consider the problem of estimating the mean parameter for a normal distribution under the $ N(\mu, 1) $ model. Table~\ref{tab: simu-1} presents the mean squared error (MSE) for the MLE and the WLE for weight function 1 at $\alpha=1.01$ and $\alpha=1.02$ for data generated from an $N(0,1)$ distribution contaminated by an $N(0,25)$ distribution. This is thus a scale contamination scenario. The level of contamination $\epsilon$ varies between $0\%$ to $50\%$, at intervals of $10\%$.
	\begin{table}[!htbp]
		\begin{center}
			\caption{Mean squared error of the proposed estimators: Scale contamination}
			\begin{tabular}{c c c c}
				\hline
				&\multicolumn{3}{ c}{Mean Squared Error}\\
				\cline{2-4}
				$\epsilon$&MLE&WLE$_{1,\alpha=1.01}$&WLE$_{1,\alpha=1.02}$\\
				\hline
				0\%     &	0.0339& 	0.0385& 	0.0434\\
				10\%    &	0.1179& 	0.0526& 	0.0577\\
				20\%	&	0.1913& 	0.0704& 	0.0711\\
				30\%	&	0.2839& 	0.1147& 	0.1045\\
				40\%	&	0.3635& 	0.1900& 	0.1587\\
				50\%	&	0.4538& 	0.2877& 	0.2379\\
				\hline
			\end{tabular}\\
			\footnotesize \textsc{Note}: the sample size for each of the cases was 30. Each mean squared error is based on 1000 replications. \footnotesize The data generating distribution is $(1-\epsilon) N(0,1)+\epsilon N(0,25)$, where $\epsilon$ is the contamination proportion.
			\label{tab: simu-1}
		\end{center}
	\end{table}
	
	Since the existence of multiple roots to the weighted likelihood estimating equations is a natural issue here, we do a bootstrap root search as proposed by~\cite{markatou98}. At each level of contamination, we picked $1000$ replicates, each of size $30$. Then we took $50$ independent bootstrap samples of size $3$ from each sample. Using the MLEs of these samples as starting values we obtained the WLEs for each such starting value and identified the unique roots. In presence of multiple nondegenerate roots, we followed the suggestion of \cite{biswas14} and picked the root for which the sum of weights is second highest, provided, the corresponding sum was at least as high as $25\%$ of the total weight. After choosing the roots, we calculated the mean squared error around $0$. The mean squared error for the WLEs are much smaller than the MLE under contamination.
	
	Next we consider a location contamination example where an $N(0,1)$ model is contaminated by an $N(5,1)$ distribution. Table~\ref{tab: simu-2} represents the mean squared errors in estimating the mean parameter. Since, the presence of multiple roots is highly likely in this case as well, we employed the exact same strategy as described in the previous paragraph and calculated the mean squared error around $0$.
	\begin{table}[!htbp]
		\begin{center}
			\caption{Mean squared error of the proposed estimators: Location contamination}
			\begin{tabular}{c c c c}
				\hline
				&\multicolumn{3}{ c}{Mean Squared Error}\\
				\cline{2-4}
				$\epsilon$&MLE&WLE$_{1,\alpha=1.01}$&WLE$_{1,\alpha=1.02}$\\
				\hline
				0\%&		0.0323& 	0.0356& 	0.0429\\
				10\%&		0.3668& 	0.0631& 	0.0526\\
				20\%&		1.1414& 	0.1487& 	0.0907\\
				30\%&		2.4672& 	0.5508& 	0.4725\\
				40\%&		4.3454& 	3.7214& 	3.4854\\
				50\%&		6.4610& 	11.0333& 	10.7086\\
				\hline
			\end{tabular}\\
			\begin{footnotesize}
			\textsc{NOTE}: the sample size for each of the cases was 30. Each mean squared error is based on 1000 replications. \footnotesize The data generating distribution is $(1-\epsilon) N(0,1)+\epsilon N(5,1)$, where $\epsilon$ is the contamination proportion.
			\end{footnotesize}
			\label{tab: simu-2}
		\end{center}
	\end{table}
 
	Under contamination, the MSE for MLE blows up. However, the WLEs perform well in identifying the target value of zero and hence produce substantially smaller MSE, at least for smaller values of contamination. However, as the contamination proportion tends to $ 50\% $, the performance of the WLE becomes poorer (in terms of increased MSE), and this phenomenon demands an explanation. At $ 50\% $ contamination, both the components of the mixture become equally strong in terms of their representation in the sample, and the final selection of the root becomes a toss up between the means of the two components. The method, therefore, chooses a root around $0$ half of the time, and a root around 5 in the remaining half. Thus the empirical mean squared error is supposed to be of the order of $ (5-0)^2/2 = 12.5 $, which is what we approximately observe. In case of the MLE, however, the process throws out an estimator which is close to the average of the two component means, which is 2.5. Thus the mean squared error in this case is of the order of $(2.5-0)^2=6.25$, close to the observed value in Table~\ref{tab: simu-2}.

	We consider the exponential distribution for our next simulation study. Smoothing based on usual kernels produces nonnegative estimated densities for part of the negative side of the real line, and more sophisticated kernels are needed for this case, complicating the theory. Such a difficulty does not arise in this case, and the estimation method can easily proceed as in the normal case. We employ the same weight function and the same scheme for choosing the root as before. We assume an exponential($\lambda$) model, $\lambda$ being the rate parameter. Table~\ref{tab: simu-3} presents the mean squared errors for the estimate of $\lambda$ when the data are generated by an exponential(1) distribution contaminated by an exponential(1/5) distribution. The WLE outperforms MLE in terms of the MSE.
	\begin{table}[!htbp]
		\begin{center}
			\caption{Mean squared error of the proposed estimators: Exponential model}
			\begin{tabular}{c c c c}
				\hline
				&\multicolumn{3}{ c}{Mean Squared Error}\\
				\cline{2-4}
				$\epsilon$&MLE&WLE$_{1,\alpha=1.01}$&WLE$_{1,\alpha=1.02}$\\
				\hline
				0\%&		0.0373 &	0.0392 & 	0.0467\\
				10\%&		0.0997 &	0.0660 & 	0.0624\\
				20\%&		0.1919 &	0.1557 & 	0.1525\\
				30\%&		0.2797 & 	0.1997 & 	0.2094\\
				40\%&		0.3563 & 	0.2974 & 	0.2637\\
				50\%&		0.4223 & 	0.3764 & 	0.3497\\
				\hline
			\end{tabular}\\
			\begin{footnotesize}
			\textsc{NOTE} : The sample size for each of the cases was 30. Each mean squared error is based on 1000 replications. \footnotesize The data generating distribution is $ (1-\epsilon)~\text{exponential}(1) + \epsilon~\text{exponential}(1/5) $, where $\epsilon$ is the contamination proportion.\\
			\end{footnotesize}
			\label{tab: simu-3}
		\end{center}
	\end{table}

	All the above calculations have been performed in a computer with Celeron (R) Dual Core processor (32 bit) with 3 GB RAM. The computation time turns out to be sub-second, even for sample sizes as large as $1000$. The execution time increases approximately linearly with the sample size $n$ in both one parameter and two parameter cases. More details on this can be found in \href{ESM1.pdf}{Online Resource 1}. Multivariate extensions, as described in Section \ref{subsec: multivar}, has complexity that is exponential in dimensionality but linear in sample size. The choice of initial value is important to the outcome as well as the runtime of the method, as is the case for almost any iterative root solving method. Increasing the level of contamination also increases the execution time as one would expect. Varying levels of contamination and several initial values had been tried for the simulation and the time taken was always found to be similar with negligible deviances from what has been reported.
 
	The optimal choice of the tuning parameter is an important practical issue. We have proposed a data based selection criterion for this purpose which is described in detail in \href{ESM1.pdf}{Online Resource 1}.
\section{Theoretical Properties of the Weighted Likelihood Estimator} \label{sec: properties}
	In this section, we present the theoretical properties of the WLEs for the case $p=1/2$. For brevity, all the proofs have been presented in \href{ESM1.pdf}{Online Resource 1}.
	\subsection{Fisher Consistency of the Weighted Likelihood Estimators} \label{subsec: FisherCon}
		Fisher consistency is an important and desirable property of an estimator. Suppose $X_1, X_2, \cdots , X_n$ represent an i.i.d. random sample from a distribution modelled by the parametric family $\{F_\theta\}$. Then the following property holds.
		\begin{lemma} \label{lemma : FishCon}
			The proposed weighted likelihood estimator is Fisher consistent.
		\end{lemma}
	\subsection{The Influence Function of the Weighted Likelihood Estimators} \label{subsec : Influence}
		Let $\mathcal{F}_\Theta=\{F_\theta : \theta \in \Theta \subset \mathbb{R}\}$ be the parametric model and let $T: \mathcal{G} \rightarrow \Theta$ be a functional from a relevant class ($ \mathcal{G} $) of distributions to the parameter space $ \Theta $. From Lemma \ref{lemma : FishCon} we know, $ T(F_\theta) = \theta $. To find the influence function of the proposed estimators, we consider the $\epsilon$-contaminated version of the true distribution function $G$ given by
		\begin{equation} \label{eq: 3rdeq}
			G_\epsilon(x)=(1-\epsilon)G(x)+\epsilon\Delta_y(x)
		\end{equation}
		where $\Delta_y(x)$ is the distribution function of $\chi_y$, the random variable which puts all its mass on $y$. Denote by $\bar{\Delta}_y(x)=\mathbb{P}(\chi_y \geqslant x)$ and $\bar{G}(x)= \mathbb{P}(X \geqslant x)$, with $X$ being the random variable having the distribution function $G$. We consider a general distribution $G$, not necessarily in the model.
		\begin{theorem} \label{thm : influence}
			The influence function of the proposed estimator is 
			\begin{equation}T^\prime(y)=\frac{\partial}{\partial\epsilon}\theta_\epsilon\bigg\rvert_{\epsilon=0}=D^{-1}N
			\end{equation}
			where $\theta_G=T(G)$, $\theta_\epsilon$ is the functional corresponding to the contaminated distribution in (\ref{eq: 3rdeq}) and
			\begin{align*}
				D=&\left[ \int_{\mathcal{X}_1} H^\prime(\tau(x))u_{\theta_G}(x)\frac{\nabla F_{\theta_G}(x)}{F_{\theta_G}(x)}(\tau(x)+1) dG(x) \right.\\
				&+~\int_{\mathcal{X}_2} H^\prime(\tau(x))u_{\theta_G}(x)\frac{\nabla S_{\theta_G}(x)}{S_{\theta_G}(x)}(\tau(x)+1) dG(x) + ~\left. \int H(\tau(x))\nabla (-u_{\theta_G}(x)) dG(x) \right],\\
				N=&~\left[ H(\tau(y))u_{\theta_G}(y) + \int_{\mathcal{X}_1} H^\prime(\tau(x))\Delta_y(x)\frac{u_{\theta_G}(x)}{F_{\theta_G}(x)} dG(x) \right.\\
				&+~\int_{\mathcal{X}_2} H^\prime(\tau(x))\bar{\Delta}_y(x)\frac{u_{\theta_G}(x)}{S_{\theta_G}(x)} dG(x) - ~\left. \int H^\prime(\tau(x))(\tau(x)+1)u_{\theta_G}(x) dG(x)\right],
			\end{align*}
			where $\tau \equiv \tau_{G,\theta_G}$, $\mathcal{X}_1 = \{x \in \mathcal{X} : F_\theta(x) \leqslant 1/2\}$ and $\mathcal{X}_2 = \{x \in \mathcal{X} : F_\theta(x) > 1/2\}$ and $\mathcal{X}$ being the support of the distribution. Note that $\mathcal{X}_1$ and $\mathcal{X}_2$ are disjoint and $\mathcal{X}=\mathcal{X}_1 \cup \mathcal{X}_2$. When the true distribution $G$ belongs to the model, i.e., $G \equiv F_\theta $ for some $ \theta \in \Theta $, then the influence function takes the simple form
			$ T^\prime(y)=\left[\int -\nabla u_\theta(x)dF_\theta\right]^{-1}u_\theta(y)=I^{-1}(\theta)u_\theta(y) $
			which is same as the influence function of the MLE.
		\end{theorem}
	\subsection{Location-Scale Equivariance} \label{subsec: equivariance}
		We consider a location-scale family characterized by either of the following equivariant formulations,
		\begin{enumerate}
			\item $f_{(\mu,\sigma)}(x)=\frac{1}{\sigma}f_{(0,1)}\left(\frac{x-\mu}{\sigma}\right)$,
			\item $F_{(\mu,\sigma)}(x)=F_{(0,1)}\left(\frac{x-\mu}{\sigma}\right)$,
		\end{enumerate}
		Let $\theta=(\mu,\sigma)$ represent our parameter of interest. Consider i.i.d. observations $ Z_1, Z_2,\cdots, Z_n $ from a location-scale family $F_{(\mu,\sigma)}$. Let $(\hat{\mu},\hat{\sigma})$ be the weighted likelihood estimate of the parameter $\theta$. Consider the transformation $ X_i = a + b Z_i,~a\in \mathbb{R},~b > 0,~i = 1, 2, \cdots, n $. Then our WLE is location-scale equivariant in the sense $(a+b\hat{\mu},b\hat{\sigma})$ is the estimated parameter vector for the transformed data.
		\begin{theorem} \label{thm : equivar}
			The proposed weighted likelihood estimators are location-scale equivariant.
		\end{theorem}
	\subsection{Consistency and Asymptotic Normality} \label{subsec: consistency}
		The consistency and asymptotic efficiency of the WLEs hold under the regularity conditions presented in \href{ESM1.pdf}{Online Resource 1}. While the general case involving multiple parameters can be handled by making the conditions more complicated and by routinely extending the proof, here we consider the case of a scalar parameter. 
		\begin{theorem} \label{thm : asymptots}
			Let the true distribution belong to the model, $\theta_0$ be the true parameter and let $\hat{\theta}_{n,WLE}$ be the weighted likelihood estimator. Under conditions (C1) - (C6), mentioned in \href{ESM1.pdf}{Online Resource 1}, the following results hold.
			\begin{enumerate}
				\item The convergence \[ \sqrt{n}\left\lvert A_n - \frac{1}{n}\sum_{i=1}^n u_{\theta_0}({X}_i)\right\rvert \rightarrow 0 \] holds in probability, where $ A_n = \frac{1}{n} \sum_{i=1}^n H(\tau_{n,\theta_0}(X_i)) u_{\theta_0}(X_i) $.
				\item The convergence \[ \left\lvert B_n - \frac{1}{n}\sum_{i=1}^n \nabla u_{\theta_0}({X}_i) \right\rvert \rightarrow 0\] holds in probability, where $B_n=\frac{1}{n}\sum_{i=1}^n \nabla (H(\tau_{n,\theta} (X_i)) u_\theta(X_i))\big\rvert_{\theta=\theta_0} $.
				\item $ C_n = O_p(1)$, where $C_n=\frac{1}{n}\sum_{i=1}^n \nabla_2 (H(\tau_n({X}_i)) u_{\theta}({X}_i))\rvert_{\theta=\theta^\prime}$. Here $\theta^\prime$ is on the line segment joining $\theta_0$ and $\hat{\theta}_{n,WLE}$ and $\nabla_2$ represents second derivative with respect to $\theta$.
			\end{enumerate}
		\end{theorem}
		Using the above results, the consistency of the WLE $\hat{\theta}_{n,WLE}$ follows from \cite{serfling80} and \cite{lehmann2006theory}. A straightforward Taylor expansion of the weighted likelihood estimating equation
		\begin{equation}
			\frac{1}{n}\sum_{i=1}^nH(\tau_{n,\hat{\theta}_{n,WLE}}({X}_i))u_{\hat{\theta}_{n,WLE}}({X}_i)=0
		\end{equation}
		around $ \theta = \theta_0 $ leads to the relation
		\begin{equation}
			\sqrt{n}(\hat{\theta}_{n,WLE} - \theta_0) = -\frac{\sqrt{n} A_n}{B_n + \frac{(\hat{\theta}_{n,WLE}-\theta_0)}{2}C_n} 
		\end{equation}
		which, together with the above results, immediately yields  \[ \sqrt{n}(\hat{\theta}_{n,WLE} - \theta_0) \xrightarrow{\mathcal{D}} Z^\ast \sim N(0,I^{-1}(\theta_0)). \]
\section{Higher Order Influence Function Analysis} \label{sec: hif}
	In Section~\ref{subsec : Influence} we have seen that the proposed weighted likelihood estimator has the same influence function as the maximum likelihood estimator. So, the influence function approach will not show the WLE to be any more robust than the MLE. But in reality, in all our studies and data analysis, the proposed estimator exhibits strong robustness properties in contrast to the MLE. This indicates the inadequacy of the influence function as a tool for measuring robustness. As the influence function represents a first order approximation, one can opt for a higher order analysis of robustness in a situation where the data are contaminated at a single point. In this section, we discuss the second order influence function analysis for our proposed estimator. If the second order bias prediction turns out to be substantially smaller than the first order prediction, it not only demonstrates the robustness of the estimator, but also indicates the inadequacy of the first order influence function approach in this case. The faster the second order approximation deviates from the first, the greater is the inadequacy of the latter.

	As in Section~\ref{subsec : Influence}, we take $G_\epsilon=(1-\epsilon)G + \epsilon\Delta_y$ to be the distribution function $G$ contaminated at a point $y$ by an infinitesimally small proportion $\epsilon$, with $\Delta_y$ being the distribution function of the random variable degenerate at $y$. $\bar{\Delta}_y$ and $\bar{G}$ are also accordingly defined. Let $T(G_\epsilon)=T((1-\epsilon)G + \epsilon\Delta_y)$ with $T$ being the functional of interest as before. The influence function of the functional $T(\cdot)$ is given by \[T^\prime(y)=\frac{\partial T(G_\epsilon)}{\partial \epsilon}\bigg\rvert_{\epsilon=0}.\]Viewed as a function of $\epsilon$, $\Delta T(\epsilon)=T(G_\epsilon)-T(G)$ quantifies the amount of bias and describes how the bias changes with contamination. The second order Taylor expansion gives
	\begin{equation}
		\label{eq: soi}
		T(G_\epsilon)-T(G) \approx \epsilon T^\prime(y)+\frac{\epsilon^2}{2}T^{\prime\prime}(y),
	\end{equation}
	where $T^{\prime\prime}(y)$ is the second derivative of $T(G_\epsilon)$ evaluated at $\epsilon=0$. We will find an expression for $T^{\prime\prime}(y)$ and then ascertain the expected behaviour of the proposed estimator with changes in the level of contamination using these expressions. The detailed computation of the terms in the general scenario can be found in \href{ESM1.pdf}{Online Resource 1}. We now demonstrate how the second order influence function analysis differs from the first order analysis. We consider an $N(1,1)$ population contaminated at $y=10$ and plot the (second order) absolute bias against the level of contamination ($\epsilon$). 
	
	Consider weight function 1. When $G \equiv F_{\theta_G}$, $\tau_{G,\theta_G}\equiv0$, implying $H(\tau_{G,\theta_G}) \equiv 1$, $H^\prime(\tau_{G,\theta_G}) \equiv 0$ and $H^{\prime\prime}(\tau_{G,\theta_G}) \equiv 1-\alpha$. Then, plugging in these values in the expression for $T^{\prime\prime}(y)$, we get, \[T^{\prime\prime}(y)=I^{-1}(\theta)\left[\int_{\mathcal{X}_1} (1-\alpha)\frac{u_\theta}{F_\theta}(\Delta_y - G)^2 dG + \int_{\mathcal{X}_2} (1-\alpha)\frac{u_\theta}{S_\theta}(\bar{\Delta}_y - \bar{G})^2 dG\right.\]\[+ 2T^\prime(y)\left( \int_{\mathcal{X}_1} (\alpha -1)u_\theta \frac{\nabla F_\theta}{F_\theta}(\Delta_y - G)dG + \int_{\mathcal{X}_2} (\alpha -1)u_\theta \frac{\nabla S_\theta}{S_\theta}(\bar{\Delta}_y - \bar{G})dG\right.\]\[\left.+ \nabla u_\theta (y) + I(\theta)\right) + (T^\prime(y))^2 \left( \int \nabla_2 u_\theta dG + \int_{\mathcal{X}_1} (1-\alpha)u_\theta \left(\frac{\nabla F_\theta}{F_\theta}\right)^2 F_\theta dG +\right.\]\[\left.\left. \int_{\mathcal{X}_2} (1-\alpha)u_\theta \left(\frac{\nabla S_\theta}{S_\theta}\right)^2 S_\theta dG\right)\right].\]
 
\begin{figure}[!htbp]
		\centering
		\includegraphics[width=0.8\linewidth, height=0.65\linewidth]{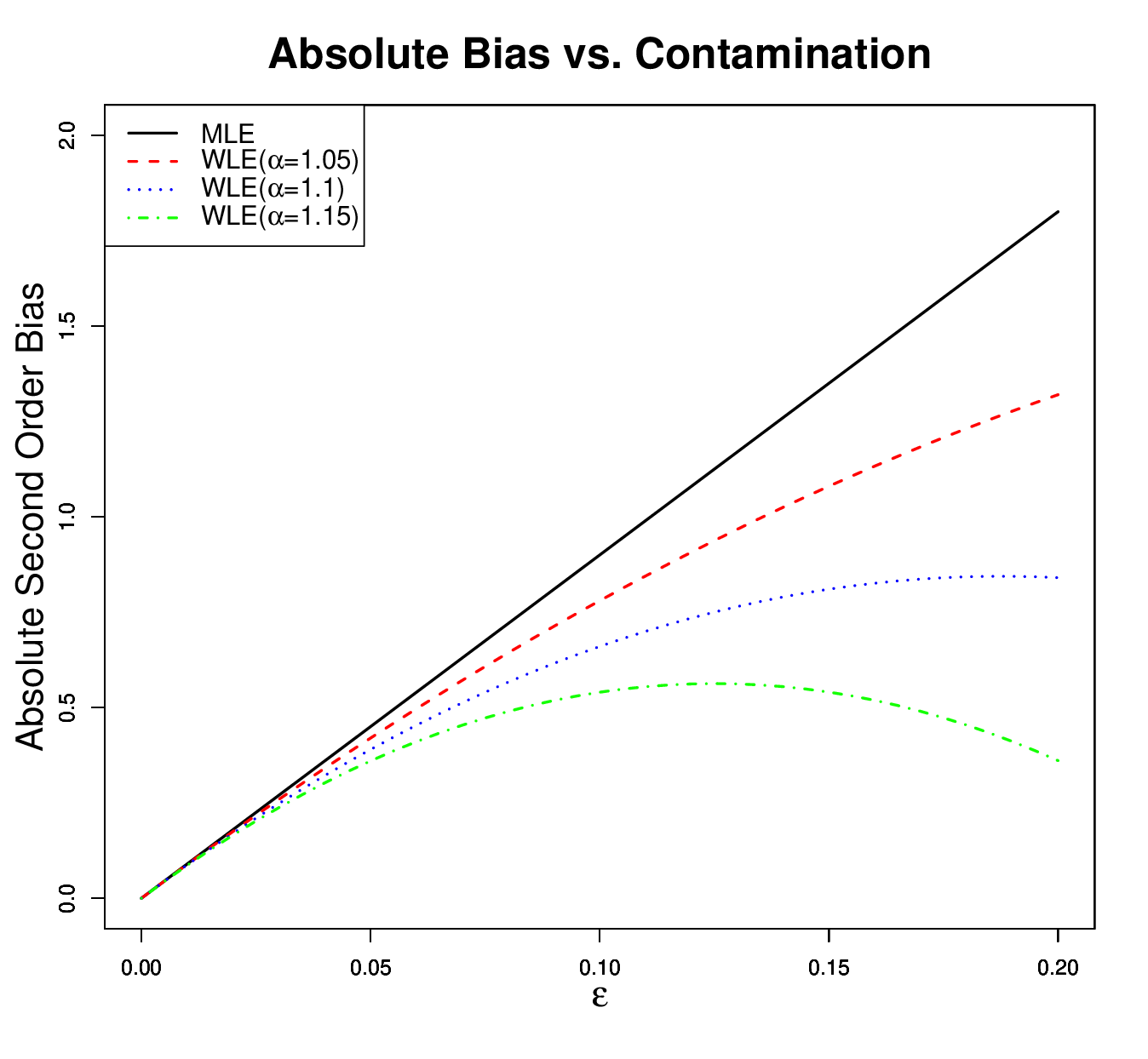} 
		\caption{The predicted, second order bias plot for the MLE and several members of the WLE class with different tuning parameters for the first weight function}
		\label{fig: hif1}
	\end{figure}

	For the MLE the first order linear approximation is exact, and the second order approximation does not alter it. But the robust WLEs, particularly those leading to sharper downweighting of large residuals lead to significantly smaller bias predictions using (\ref{eq: soi}). This is demonstrated in Figure \ref{fig: hif1}. 
\section{Extending the Method to Other Scenarios} \label{sec: extension}
	\subsection{Extension to Bivariate Scenario} \label{subsec: multivar}
		\subsubsection{Method of Estimation}
			Here we define the residual function for the proposed weighted likelihood estimation in multivariate, in particular bivariate, data set up. Unlike the univariate case where it suffices to consider the two tails of the distribution, the bivariate situation is a bit tricky as there is no such specific concept of direction to which we evaluate a tail probability. Also, basing the analysis solely on the distribution and survival functions will not faithfully reflect the position of an observation with respect to the majority of the data cloud. For this, we devise the residual function by considering probabilities of all the four quadrants for each data point. First we set up a few definitions in the spirit of the univariate case.

			Let $ (X_1, Y_1), (X_2,Y_2), \cdots, (X_n,Y_n) $ be paired i.i.d. observations from a bivariate population. Define the probabilities
			\begin{align*}
				P_{\text{ll},n}(x,y) = \frac{1}{n}\sum_{i=1}^{n} \mathbf{1}_{\lbrace X_i \leqslant x, Y_i \leqslant y\rbrace}, & ~~ P_{\text{ll},\theta}(x,y) = \mathbb{P}_\theta(X \leqslant x, Y \leqslant y), \\
				P_{\text{lg},n}(x,y) = \frac{1}{n}\sum_{i=1}^{n} \mathbf{1}_{\lbrace X_i \leqslant x, Y_i \geqslant y\rbrace}, & ~~ P_{\text{lg},\theta}(x,y) = \mathbb{P}_\theta(X \leqslant x, Y \geqslant y), \\
				P_{\text{gl},n}(x,y) = \frac{1}{n}\sum_{i=1}^{n} \mathbf{1}_{\lbrace X_i \geqslant x, Y_i \leqslant y\rbrace}, & ~~ P_{\text{gl},\theta}(x,y) = \mathbb{P}_\theta(X \geqslant x, Y \leqslant y), \\
				P_{\text{gg},n}(x,y) = \frac{1}{n}\sum_{i=1}^{n} \mathbf{1}_{\lbrace X_i \geqslant x, Y_i \geqslant y\rbrace}, & ~~ P_{\text{gg},\theta}(x,y) = \mathbb{P}_\theta(X \geqslant x, Y \geqslant y).
			\end{align*}
			Under these notations, the residual function is defined as
			\begin{align} \label{ResidualBi}
			\tau_{n,\theta}(x,y) = 
			\begin{cases}
				\dfrac{P_{\text{ll},n}(x,y)}{P_{\text{ll},\theta}(x,y)} - 1, & \text{if } ~ P_{\text{ll},\theta} = \underset{i,j}{\text{min}} \{ P_{ij,\theta} \} \\
				\dfrac{P_{\text{lg},n}(x,y)}{P_{\text{lg},\theta}(x,y)} - 1, & \text{if } ~ P_{\text{lg},\theta} = \underset{i,j}{\text{min}} \{ P_{ij,\theta} \} \\
				\dfrac{P_{\text{gl},n}(x,y)}{P_{\text{gl},\theta}(x,y)} - 1, & \text{if } ~ P_{\text{gl},\theta} = \underset{i,j}{\text{min}} \{ P_{ij,\theta} \} \\
				\dfrac{P_{\text{gg},n}(x,y)}{P_{\text{gg},\theta}(x,y)} - 1, & \text{if } ~ P_{\text{gg},\theta} = \underset{i,j}{\text{min}} \{ P_{ij,\theta} \},
			\end{cases}
			\end{align}
			where $P_{ij,\theta} \equiv P_{ij,\theta} (x,y) \,,\ i,j \in \lbrace \text{l,g}\rbrace $. Then the weight $ w_\theta(x, y) $ is constructed exactly as in the univariate case. With $ u_\theta(x,y) $ representing the bivariate score function, our estimator of the parameter $ \theta $ is now obtained as the solution of the equation
			\begin{equation}\label{eq:BivariateWLEE}
				\sum_{i=1}^{n} w_\theta(X_i,Y_i)u_\theta(X_i,Y_i) = 0.
			\end{equation}
			
			The definition of the residual function above can similarly be extended for data with dimension $ d > 2 $. In that case, we have to take the associated $ 2^d $ orthant probabilities into consideration.
		\subsubsection{Influence Function in the Bivariate Situation} \label{subsubsec:BivariateInfluenceFunction}
			Under the mathematical set up described above, we set out to evaluate the influence function of the WLE. For this, we first define the residual function under the true model as
			\begin{equation}\label{eq:ResidualBiTrue}
				\tau_{G,\theta}(x,y) =
				\begin{cases}
					\frac{G_{\text{ll}}(x,y)}{P_{\text{ll},\theta}(x,y)} - 1, & \text{if } ~ (x,y) \in S_{\text{ll}}(\theta) \\
					\frac{G_{\text{lg}}(x,y)}{P_{\text{lg},\theta}(x,y)} - 1, & \text{if } ~ (x,y) \in S_{\text{lg}}(\theta) \\
					\frac{G_{\text{gl}}(x,y)}{P_{\text{gl},\theta}(x,y)} - 1, & \text{if } ~ (x,y) \in S_{\text{gl}}(\theta) \\
					\frac{G_{\text{gg}}(x,y)}{P_{\text{gg},\theta}(x,y)} - 1, & \text{if } ~ (x,y) \in S_{\text{gg}}(\theta),
				\end{cases}
			\end{equation}
			where $ S_{ij}(\theta) = \left\lbrace (x,y): P_{ij,\theta}(x,y) = \underset{r,s\in\lbrace \text{l,g}\rbrace}{\min}~ P_{rs,\theta}(x,y)\right\rbrace,~~i,j \in \lbrace \text{l,g}\rbrace $. On the basis of these, we present the influence function as below.
			\begin{theorem}[Influence Function in Bivariate Set Up]
				Under the assumption
				\begin{align}\label{eq:EETerm2BigOEpsilon}
						\int\int \lbrace H(\tau_{G_\epsilon,\theta_\epsilon}(x,y)) u_{\theta_\epsilon}(x,y) -  H(\tau_{G,\theta_G}(x,y)) u_{\theta_G}(x,y) \rbrace 
						 \times \mathbf{1}_{\lbrace S_{ij}(\theta_\epsilon) \cap S_{rs}(\theta_G)\rbrace}(x,y)dG(x,y) = O(\epsilon)
				\end{align}
				for every pair $ (r,s) \neq (i,j) $, the influence function of the weighted likelihood estimator in the bivariate set up is given by
				\begin{equation}\label{eqC6:IFofBivariateWLE}
					IF((x_o,y_o),T,G) = \text{D}^{-1}\text{N},
				\end{equation}
				where $ \text{D} $ and $ \text{N} $ are as defined in \href{supplementary.pdf}{Online Resource 1}.
				
				In case the true distribution belongs to the model, i.e., $ G \equiv F_{\theta_0} $, it follows that $ \tau_{G,\theta}(x,y) \equiv 0 $ which in turn implies $ H(\tau_{G,\theta}(x,y)) \equiv 0~\text{and}~H^\prime(\tau_{G,\theta}(x,y)) \equiv 0 $. The influence function in that situation reduces to
				\begin{equation}\label{eqC6:IFofBivariateWLEAtModel}
					IF((x_o,y_o),T,F_{\theta_0}) = I^{-1}(\theta_0)u_{\theta_0}(x_o,y_o),
				\end{equation}
				which is the influence function of the MLE. Here $ I(\theta_0) $ is the bivariate Fisher information matrix in this set up.
			\end{theorem}
			It is also to be noted that while the result is presented in case of bivariate data only, the proof (as provided in \href{supplementary.pdf}{Online Resource 1}) is more general in that it is applicable for data of dimensions higher than two.
		\subsubsection{A Bivariate Example} \label{subsec:BivariateIllustration}
\begin{figure}[!htbp]
		\begin{center}
			\includegraphics[height=0.65\linewidth,width=0.8\linewidth]{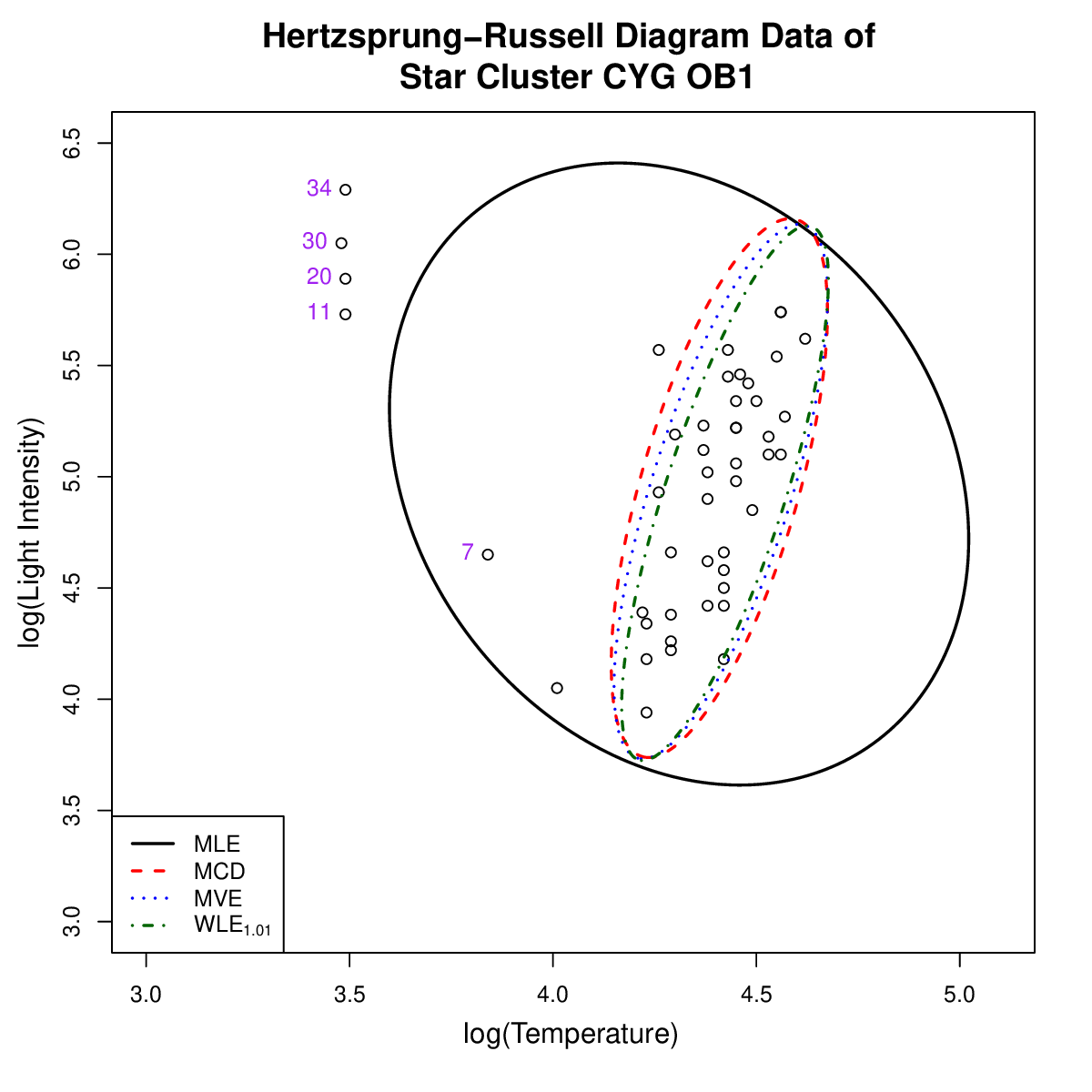}
			\caption{The 95\% concentration ellipses for MLE, MCD, MVE and the proposed WLE at $ \alpha = 1.01 $ for the Hertzsprung-Russell data set}
			\label{fig: HRplot}
		\end{center}
	\end{figure}

			We use the Hertzsprung-Russell data set \citep{russo87} which contains observations for the luminosity of stars versus their effective temperature in the logarithmic scale. There are four large outliers in the upper left hand corner of Figure~\ref{fig: HRplot}, as well as a few minor outliers. We treat these data as a sample from a bivariate normal population $ BN(\mu_1,\mu_2,\sigma_1^2,\sigma_2^2,\rho) $.
							
			The parameter estimates obtained by the maximum likelihood and weighted likelihood estimation methods are displayed in Table \ref{tab: HREstimates}. Clearly there is a major change in the scale of the log-temperature variable as well as in the covariance component. To visually demonstrate the effect of weighting the data points, we have presented the 95\% concentration ellipses based on the parameter values obtained by the MLE and the WLE under the tuning parameter $ 1.01 $. To compare the performance of our method, we also add the concentration ellipses corresponding to two well-known multivariate location and scatter estimates, viz., the minimum covariance determinant (MCD) and the minimum volume ellipsoid (MVE) estimates. All these are combined and presented in Figure~\ref{fig: HRplot}. The MCD and MVE estimates of the five parameters are also presented in Table~\ref{tab: HREstimates}.
			\begin{table}[!htbp]
				\begin{center}
					\caption{Estimates obtained for the Hertzsprung-Russell data set}
					\begin{tabular}{c c c c c c}
						\hline
						Method & $ \hat{\mu}_1 $ & $ \hat{\mu}_2 $ & $ \hat{\sigma}_1^2 $ & $ \hat{\sigma}_2^2 $ & $ \hat{\rho} $ \\
						\hline
						MLE & $4.3100$ & $5.0121$ & $0.0846$ & $0.3263$ & $-0.2104$\\
						MCD & $4.4090$ & $4.9490$ & $0.0118$ & $0.2449$ & $0.6548$\\
						MVE & $4.4127$ & $4.9335$ & $0.0115$ & $0.2410$ & $0.7313$\\
						WLE$ _{1.01} $ & $4.4222$ & $4.9264$ & $0.0111$ & $0.2479$ & $0.7919$\\
						\hline
					\end{tabular}
					\label{tab: HREstimates}
				\end{center}
			\end{table}
							
			It is evident that the weighted likelihood scheme, like the two other existing robust estimators, produces a much more meaningful concentration ellipse, covering the majority of the observed data and sacrificing the outlying points. The outlier resistant property of the WLEs causes the corresponding ellipses to shrink sharply compared to the highly liberal and practically useless ellipse produced by maximum likelihood.
			
			We have also applied the proposed extension of the weighted likelihood method on the beetle data set from \cite{lubischew62} to uncover distinct populations from the data. However, similar exercise has been performed in \cite{biswas14} with another bivariate data set. Hence, we relegate this analysis to \href{ESM1.pdf}{Online Resource 1}.
	\subsection{Normal Linear Regression} \label{subsec: regression}
		\subsubsection{Methodology} \label{subsubsec: regress_est}
			Let us consider a homoscedastic linear regression model $ Y_i=\beta_0+\beta_1x_i+\epsilon_i,~i=1,2,\cdots,n. $ We also assume that the errors are independent and $\epsilon_i \sim N(0,\sigma^2)$ for all $i=1,2,\cdots,n.$ Now, since $\epsilon_i=Y_i-\beta_0-\beta_1x_i$, we have $\left(Y_i-\beta_0-\beta_1x_i\right)/\sigma \sim N(0,1)$. Let ${Z}_i=(Y_i-\beta_0-\beta_1x_i)/\sigma \,\ i=1,2,\cdots,n.$ We define \[F_{n,\theta}(z)=\frac{1}{n}\sum_{i=1}^n I({Z}_i \leqslant z) \,,\ S_{n,\theta}(z)=\frac{1}{n}\sum_{i=1}^n I(Z_i \geqslant z)\] where $\theta=(\beta_0,\beta_1,\sigma)$. If the true value of the parameters are specified, $F_{n,\theta}(z)$ converges to $\Phi(z)$, the standard normal cumulative distribution function at $z$. With this, we now define 
			\begin{equation} \label{eq:RegressionResidualStd}
				\tau_{n,\theta}(z) =
				\begin{cases}
					\frac{F_{n,\theta}(z)}{\Phi(z)} - 1,& \text{if } 0 < \Phi(z) \leqslant p,\\		
					\qquad 0,& \text{if } p < \Phi(z) < 1-p,\\
					\frac{S_{n,\theta}(z)}{1-\Phi(z)} - 1& \text{if } 1-p \leqslant \Phi(z) < 1.\\
				\end{cases}
			\end{equation}
			We define the weight $ w_\theta(z) = H(\tau_{n,\theta}(z)) $ similar to the univariate case and obtain the weighted likelihood estimates of $\theta=(\beta_0,\beta_1,\sigma)$ by solving the weighted likelihood score equation
			\begin{equation}\label{eq:RegressionWLEE}
				\frac{1}{n}\sum_{i=1}^n w_\theta(Z_i)u_{\theta}(Z_i)=0.
			\end{equation}
			The method can obviously be extended to the linear multiple regression case in a routine manner.
		\subsubsection{Influence Function} \label{subsubsec:RegressionIF}
			Here, we evaluate the robustness of the weighted least squares (WLS) estimator by calculating the first-order influence function.
			\begin{theorem}
				The influence function of the weighted least squares functional is given by
				\begin{equation}\label{eq:IFofWLS}
					IF_{\btta}((\bx_o, y_o), T, G) = D_{\btta_G,\text{W}}^{-1}(G) N_{\btta_G,\text{W}}(G),
				\end{equation}
				where $ D_{\btta_G,\text{W}}(G) $ and $ N_{\btta_G,\text{W}}(G) $ are as defined in \href{supplementary.pdf}{Online Resource 1}. In case the error distribution is normal, the influence function of the WLS functional reduces to that of the LS functional.
			\end{theorem}
		\subsubsection{Example} \label{subsubsec: regress_exmpl}
			To illustrate the use of the WLE in the regression scenario, we use the Animals data~\citep{russo87}. These data consist of 28 observations of the body weights and brain weights of different land animals. The model used for regression is\[\log Y_i=\beta_0 + \beta_1\log x_i + \epsilon_i\]where $Y_i$ and $x_i$ are brain and body weights of the $i$th animal respectively and $\epsilon_i \sim N(0,\sigma^2)$. Table~\ref{tab: animals-comp} contains the estimated values of the regression parameters obtained from the ordinary least squares method (which are the ML estimates under normality of errors) and the ones obtained from the weighted likelihood estimation method.
			\begin{table}[!htbp]
				\begin{center}
					\caption{Estimates obtained for the Animals data}
					\begin{tabular}{c c c c}
						\hline
						Method&$\hat{\beta_0}$&$\hat{\beta_1}$&$\hat{\sigma}$\\
						\hline
						MLE&2.5549&0.4960&1.5320\\
						WLE$_{1;\alpha=1.05}$&1.8054&0.7673&0.3125\\
						\hline
					\end{tabular}
					\label{tab: animals-comp}
				\end{center}
			\end{table}
			
\begin{figure}[!htbp]
		\centering
		\includegraphics[width=0.8\linewidth, height=0.65\linewidth]{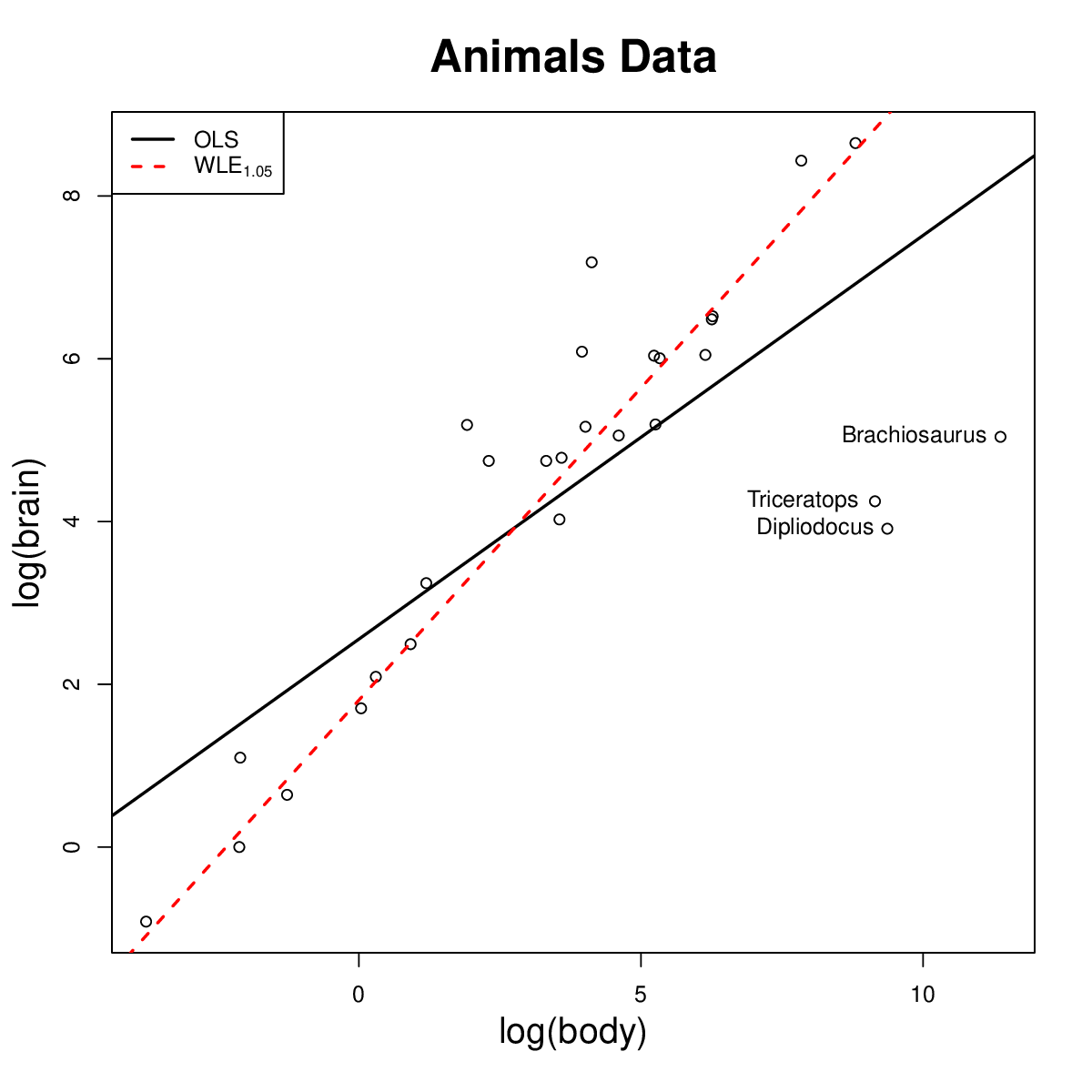}  
		\caption{The scatter plot of the animals data and the OLS and weighted likelihood regression lines}
		\label{fig: reg1}
	\end{figure}

			Figure~\ref{fig: reg1} displays different regression lines together with the scatter plot of the data. Clearly, the weighted likelihood method keeps the effect of the outliers in check unlike the ordinary least squares (OLS) estimators. Note that the scale estimate reported with the least squares is the MLE, and not the unbiased estimator.
		\subsubsection{Choice of Initial Value} \label{subsubsec: init_regress}
\begin{figure}[!htbp]
		\centering
		\includegraphics[width=0.8\linewidth, height=0.65\linewidth]{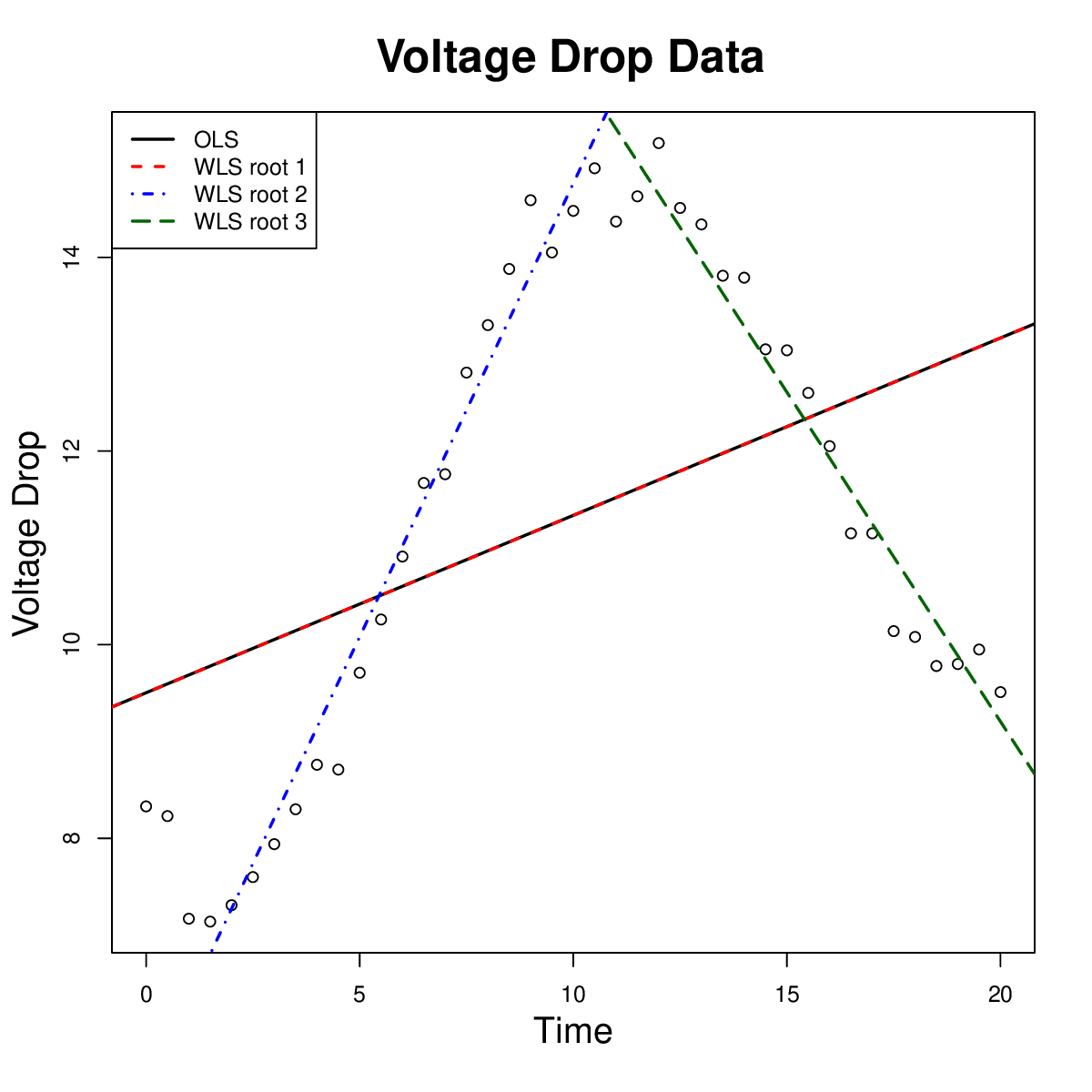} 
		\caption{Voltage Drop Data}
		\label{fig: reg2}
	\end{figure}

			As have been observed in several numerical exercises, the choice of the initial value for the estimating equation plays an important role in determining the estimate in simple estimation scenarios. For regression analysis also, the same is true. In a mixed population, where the components are significantly different, the different choices of initial values lead to different values of the regression estimate. As an example we use the voltage drop data set \citep{monty12} which has 41 observations as depicted in the scatter plot of Figure~\ref{fig: reg2}.

			Clearly, the data cannot be appropriately  modeled by a single regression line. The OLS line passes through the center of the scatter plot without providing a meaningful description of the data. However, our weighted likelihood procedure, applied with the first weight function and $\alpha = 1.02$, clearly identifies three roots. While one root is essentially an MLE like root, the other two indicate that very different regressions are appropriate for the first and the second part of the data. The coefficients are presented in Table~\ref{tab: voltage-comp}, and the fitted lines are displayed in Figure~\ref{fig: reg2}.
			\begin{table}[!htbp]
				\begin{center}
				\caption{The different roots for the Voltage drop data}
				\begin{tabular}{c c c c}
					\hline
					Method&$\hat{\beta}_0$&$\hat{\beta}_1$&$\hat{\sigma}$\\
					\hline
					MLE& 9.4855&0.1860&2.3301\\
					WLS root 1&9.5031&0.1832&2.2869\\
					WLS root 2&5.3997&0.9364&0.4142\\
					WLS root 3&22.8118&$-0.6803$&0.4233\\
					\hline
				\end{tabular}
				\label{tab: voltage-comp}
				\end{center}
			\end{table}
\section{Conclusion} \label{sec: conclusion}
	Our work comes along the route of the statistical community's efforts to estimate parameters robustly and efficiently, and is a natural sequel to the works of \citet{lindsay94}, \citet{field1994robust} and \citet{markatou98}. While Lindsay's proposal of distance based approach is very useful in discrete case, in continuous case, there are obvious problems which were later partially resolved in \cite{markatou98}. However, that method involves appropriate nonparametric smoothing methods and so it still has to deal with bandwidth selection and other problems; models with bounded support could be an irritant. The weighted likelihood estimation approach discussed in this paper provides a simple solution to such problems.

	Although we have mainly focused on normally distributed models in the continuous cases, we have provided a real data example and a simulation study in other models such as the exponential. We have demonstrated its extension to bivariate problems, as well as to the regression situation. We have briefly explored the application of the method in three dimensions, and indicated the extensions to higher dimensions. Many other scenarios where the methods of Agostinelli and colleagues have found application, such as censored likelihood estimation or Bayesian inference would be accessible to our method. On the whole we expect that it will be a useful tool for the practitioners.
	 
	In this manuscript, all the numerical results presented are with respect to $ p = 0.5 $. We have also experimented with several smaller values of $p$. We have observed that there is no dramatic difference between the estimates over the different values of $p$. This is partially because in all our numerical examples the proportion of outliers is relatively small so that relatively small values of $p$ are able to cover all the outliers. Some additional numerical results involving values of $p < 0.5$ are presented in \href{ESM1.pdf}{Online Resource 1}.
	 
	In this paper, we have focused quite extensively on the root selection issue; another problem, is the issue of the tuning parameter selection. We believe this issue is too important to receive a cursory or peripheral treatment and needs to be studied on its own as a separate problem. A data based suggestion for tuning parameter selection is presented in \href{ESM1.pdf}{Online Resource 1}. However, for a quick recommendation we propose, on the basis of repeated simulations, the values $\alpha = 1.01$,  $k = 1.01$, $\xi = 10$ and $ (d_1, d_2) = (2.1, 1) $ for the respective weight functions $1,2,3$ and $4$. As all the WLEs are first order efficient, the choice of the tuning parameter does not determine the asymptotic efficiency of the estimator.
 
\section*{Declaration}
	The authors of this article declare no conflict of interests in publishing this article.

\section*{Figures} 

	
\clearpage
\bibliographystyle{spbasic}
\bibliography{newwle}

\end{document}